\documentclass[prx, twocolumn, superscriptaddress, notitlepage, floatfix]{revtex4-1}
\usepackage{amsmath,amsfonts, amssymb, amsthm, dsfont}
\usepackage{yfonts}
\usepackage{bm}
\usepackage{mathrsfs}
\usepackage{graphicx}
\usepackage{verbatim}
\usepackage{hyperref}
\usepackage{wasysym}
\usepackage{xcolor}
\usepackage{tikz}

\newcommand{\ket}[1]{|#1\rangle}

\renewcommand{\ol}[1]{\overline{#1}}
\newcommand{\comments}[1]{}
\newcommand{\mb}[1]{\mathbf{#1}}
\renewcommand{\cal}[1]{\mathcal{#1}}

\newcommand{\Eq}[1]{Eq.~\eqref{#1}}

\newcommand{\Fig}[1]{Fig.~\ref{#1}}

\def\Z{\mathbb{Z}}

\def\id{\mathds{1}}

\DeclareMathOperator{\Tr}{Tr}

\makeatletter
\def\l@subsubsection#1#2{}
\makeatother

\begin{document}

\title{Topological Disorder Parameter}
\author{Bin-Bin Chen}
\affiliation{
Department of Physics and HKU-UCAS Joint Institute of Theoretical and Computational Physics,
The University of Hong Kong, Pokfulam Road, Hong Kong SAR, China
}
\author{Hong-Hao Tu}
\affiliation{Institut f\"ur Theoretische Physik, Technische Universit\"at Dresden, 01062 Dresden, Germany}
\author{Zi Yang Meng}
\email{zymeng@hku.hk}
\affiliation{
Department of Physics and HKU-UCAS Joint Institute of Theoretical and Computational Physics,
The University of Hong Kong, Pokfulam Road, Hong Kong SAR, China
}
\author{Meng Cheng}
\email{m.cheng@yale.edu}
\affiliation{Department of Physics, Yale University, New Haven, CT 06520-8120, U.S.A}

\begin{abstract}
	We introduce a many-body topological invariant, called the topological disorder parameter (TDP), to characterize gapped quantum phases with global internal symmetry in (2+1)d. TDP is defined as the constant correction that appears in the ground state expectation value of a partial symmetry transformation applied to a connected spatial region $M$, the absolute value of which scales generically as $\exp(-\alpha l+\gamma)$ where $l$ is the perimeter of $M$ and $\gamma$ is the TDP. Motivated by a topological quantum field theory interpretation of the operator, we show that $e^\gamma$ can be related to the quantum dimension of the symmetry defect, and provide a general formula for $\gamma$ when the entanglement Hamiltonian of the topological phase can be described by a (1+1)d conformal field theory (CFT).  A special case of TDP is equivalent to the topological R\'enyi entanglement entropy when the symmetry is the cyclic permutation of the replica of the gapped phase. We then investigate several examples of lattice models of topological phases, both analytically and numerically, in particular when the assumption of having a CFT edge theory is not satisfied. We also consider an example of partial translation symmetry in Wen's plaquette model and show that the result can be understood using the edge CFT. Our results establish a new tool to detect quantum topological order.
\end{abstract}

\date{\today}

\maketitle

\section{Introduction}
Spontaneous symmetry breaking in many-body systems is characterized by long-range correlation of a local order parameter. On the other hand, in symmetric phases, order parameters usually have only short-range or algebraic correlations. In order to characterize the symmetric phases,  it has been proven fruitful to instead consider a family of non-local observables called the disorder operator~\cite{Kadanoff1971,Fradkin2017,Nussinov2006,Nussinov2009},  which is the symmetry transformation applied only to a certain region $M$ of the system. The ground state expectation value of the disorder operator, which will be called the disorder parameter, decays exponentially with the volume of $\partial M$, the boundary of $M$. Such a scaling behavior is characteristic of a symmetry-preserving ground state and the scaling coefficient is controlled by non-universal, short-distance details of the ground state. The subleading corrections are however often more interesting and can give rise to new universal quantities~\cite{Song2012}. For instance, recently it was observed that at (2+1)d quantum critical points, the disorder parameter can exhibit a logarithmic subleading correction~\cite{Wu2021,Wu2021b,Wang2021,Wang2021scaling,Zhao2021,Estienne2022}, whose coefficient is a universal function of opening angles of corners on the boundary of the region, generalizing similar results for entanglement entropy~\cite{Fradkin2006,Casini2006,Laflorencie2016}.  Thus the disorder parameter can provide new ways to probe the nature of a many-body wavefunction. Moreover, along with the entanglement entropy, the subleading corrections of disorder operator also provide new insights in exotic deconfined quantum critical points beyond unitary CFT~\cite{Wang2021scaling,JRZhao2022}.

In contrast, when the ground state is gapped and only contains short-range correlations, there are no subleading logarithmic corrections. In this case, the expectation is that the remaining subleading correction is a universal constant, analogous to the topological correction to the entanglement area law~\cite{KitaevPreskill2006,Levin2006,Isakov2011,Block2020,Jiang2012,JRZhao2022QiuKu}. The major question we will address in this work is the physical meaning of this universal constant.   More concretely, for an element $g$ of the symmetry group $G$, denote the corresponding disorder operator in the region $M$ by $U_M(g)$. Then we expect
\begin{equation}
    \ln  | \langle U_M(g)\rangle |=- \alpha |\partial M| + \gamma_g + \cdots.
    \label{eqn:tdp}
\end{equation}
Here $\langle\cdot \rangle$ is the ground state expectation value. $\alpha$ is a non-universal constant, and $\gamma_g\geq 0$ is the universal term that we are interested in, which will be called as the topological disorder parameter (TDP). Similar observables have been studied for point-group symmetry in fermionic topological insulators and superconductors~\cite{ShiozakiPRB2017}. Our main result in this paper is that the subleading correction $\gamma_g$ is related to the quantum dimension of the symmetry defects. As will be demonstrated below for a large class of topological phases with CFT entanglement Hamiltonian (e.g. chiral topological phase) we have $\gamma_g=\ln d_g $ where $d_g$ is the quantum dimensions of defects. Intuitively, the appearance of quantum dimension can be understood as follows: one can think of the disorder operator $U_M(g)$ as the process of creating a pair of $g$ and $g^{-1}$ symmetry defects, moving the $g$ defect along the boundary $\partial M$ and then annihilating the pair. In a pure topological theory, the amplitude of such a process is given by the quantum dimension of the defect~\cite{SET}. Here we show that this intuition is basically correct in the more generic situation, with an important subtlety that there are in general multiple distinct types of defects, which can all contribute to the disorder parameter.  We further demonstrate through examples that similar relations hold even when the entanglement Hamiltonian can not be approximated by CFTs.

The form Eq. \eqref{eqn:tdp} is clearly reminiscent of the area law for quantum entanglement. This is not a coincidence. The $n$-th R\'enyi entropy of a quantum many-body system can be regarded as basically the disorder parameter for the cyclic permutation symmetry in the replicated system~\cite{Casini2006,Casini2011,Zhao2021}. In (2+1)d gapped phases, we will show below that $\gamma_g$ in this case is equal to the well-known topological R\'enyi entanglement entropy~\cite{Levin2006,KitaevPreskill2006, FlammiaPRL2009}.

The paper is organized in the following structure. To set the stage, we first provide a short review of symmetry defects in topological phases in Sec.~\ref{Sec:SET}. Then in Sec.~\ref{Sec:III}, derivations of Eq.~\eqref{eqn:tdp} for topological phases with CFT entanglement spectra are given. In Sec.~\ref{Sec:IV}, both analytical and numerical results for chiral and non-chiral topological phases are presented. Using large-scale density matrix renormalization group (DMRG)~\cite{White1993} we compute TDPs in lattice models including the $\Z_N$ toric code and Wen's plaquette models both with transverse and longitudinal fields so the model is away from the exactly solvable limit. The obtained finite-size scaling results of TDP are consistent with the prediction of quantum dimension of the symmetry defect. Finally, Sec.~\ref{Sec:V} presents the discussion of few immediate directions.

\section{Symmetry defects in topological phases}
\label{Sec:SET}
First we briefly review the general theory of anyons and defects in (2+1)d gapped phases, following Ref. [\onlinecite{SET}] (see also [\onlinecite{TarantinoSET2016}] and [\onlinecite{TeoSET2015}] for related discussions).

In a gapped phase, quasi-particle excitations can be classified into different superselection sectors, called anyon types and labeled by $a,b,c,\dots$. We will sometimes denote the full set of labels by $\mathcal{C}$. The anyons have nontrivial exchange and braiding statistics, which completely characterize the topological order in the bulk. In particular, for each anyon type $a$ we denote by $\theta_a$ the topological twist factor, or the self exchange statistics. For a pair of anyons $a$ and $b$, the S matrix element $S_{ab}$ characterizes the mutual braiding statistics. Note that a local excitation (i.e. which can be created by a local operator) corresponds to the trivial anyon type $0$, with $\theta_0=1$ and $S_{0a}=S_{a0}=\frac{d_a}{\mathcal{D}}$. Here $d_a\geq 1$ is the quantum dimension of the anyon type $a$. If there are well-separated $n$ anyons all of type $a$ in the system, then there are asymptotically $\sim d_a^n$ number of degenerate states. $\mathcal{D}=\sqrt{\sum_{a\in \mathcal{C}}d_a^2}$ is called the total quantum dimension. The data $S_{ab}$ and $\theta_a$ satisfy a number of compatibility conditions, and can be considered as a set of topological invariants that characterize the topological order.

\begin{figure}[t]
		\centering
		\includegraphics[width=\columnwidth]{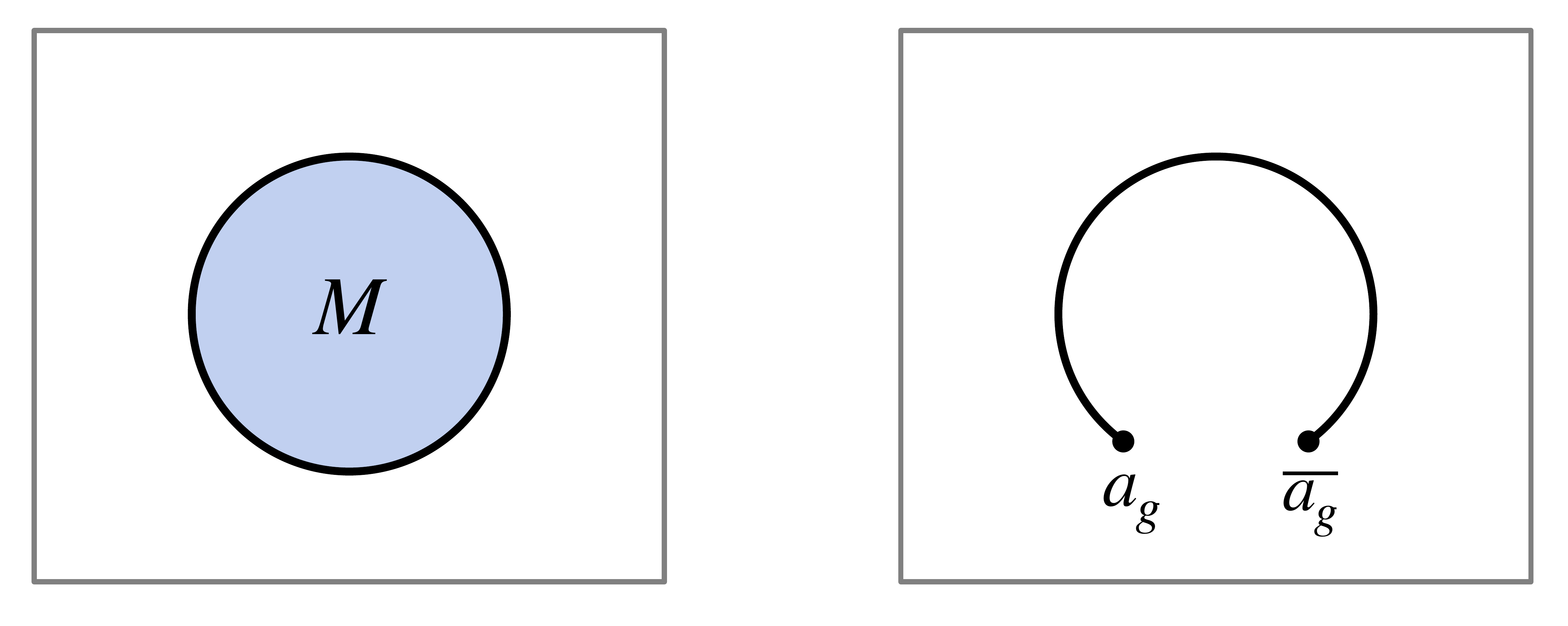}
		\caption{Left: the disorder operator for $g$ transformation on the region $M$. Right: the boundary of the disorder operator can be ``cut'' to create a pair of $g$ defects, denoted by $a_g$ and $\ol{a_g}$.}
		\label{fig:defect}
	\end{figure}

Now suppose that the underlying system has a global symmetry $g$. The symmetry can act on anyons, transforming an excitation of type $a$ into one of type $\varphi(a)$, which may be different from $a$. It has proven to be extremely useful in the theory of symmetry-enriched topological phase to introduce symmetry defects that carry $g$ fluxes. To explain this concept, first we define a disorder operator for $g$: as shown in the left panel in Fig.~\ref{fig:defect}, for a given region $M$, $U_M(g)$ is the $g$ transformation applied only to the region $M$. In a lattice model, suppose that $U(g)$ is an on-site symmetry of the form 
 \begin{equation}
		U(g)=\prod_{\mb{r}}U_\mb{r}(g),
    \end{equation}
	where $U_\mb{r}(g)$ is a unitary transformation acting on the degrees of freedom at site $\mb{r}$. Then $U_M(g)$ is given by
	\begin{equation}
		U_M(g)=\prod_{\mb{r}\in M}U_\mb{r}(g).
	\end{equation}
	Under the partial symmetry transformation the Hamiltonian becomes $H'=U_M(g)HU_M^\dag(g)$. Hamiltonian terms that are entirely supported on $M$ or the complement of $M$ do not change under the $U_M(g)$ action. Thus $H'$ only differ from $H$ along the boundary $\partial M$, where $U_M(g)$ can modify the Hamiltonian terms near the boundary nontrivially. We say that $H'$ has a $g$ defect line along $\partial M$. Equivalently, we can say that the disorder operator $U_M(g)$ creates a defect loop in the system.

Now imagine that the defect loop is cut open, i.e. there are two end points joint by a defect line. We call the end points symmetry defects labeled by $g$ and $g^{-1}$. Such a configuration can not be created by applications of disorder operators, but one can still modify the Hamiltonian along the defect line in the same way that a disorder operator would do to create open defect lines. The defining feature of a symmetry defect is that when a particle is transported around the defect, a $g$ symmetry transformation is enacted on the particle, a generalization of the Aharonov-Bohm effect.

    An important remark is in order: fixing the symmetry transformation $g$, there is a well-defined prescription for creating the defect line as explained in the previous paragraph. However, the prescription becomes ambiguous near the end points. As a result, there are actually distinct ``superselection sectors'' of $g$ defects, which only differ in the local profile at the defect point but with the same defect line and $g$ action. We denote these different types of $g$ defects by $a_g$, as shown in the right panel in Fig.~\ref{fig:defect}. The collection of all the types of $g$ defects will be denoted by $\mathcal{C}_g$. These different types of defects correspond to different ways to modify the disorder operator along the boundary.

    It turns out that the symmetry defects, while being extrinsic objects, behave in many ways like anyons. In particular, one can fuse defects to create new defects.  Each type of $g$ defect $a_g$ is associated with a quantum dimension $d_{a_g}$, which characterizes the possible exponential growth of the ground state degeneracy when multiple $a_g$ defects are present in the system. We refer the interested readers to Ref. [\onlinecite{SET}] for a more comprehensive treatment of the algebraic theory of symmetry defects.

    As shown in Ref. [\onlinecite{SET}], if the symmetry $g$ does not change anyon types, i.e. $\varphi(a)=a$ for all $a\in \mathcal{C}$, then there is at least one defect, which will be called $0_g$, with quantum dimension $d_{0_g}=1$. In other words, it is an Abelian defect. In this case, all other defects can be obtained by fusing $0_g$ with anyons: $a_g=0_g\times a$ with $a\in \mathcal{C}$, and $d_{a_g}=d_a$. If, however, the symmetry permutes anyon types, then all defects must be non-Abelian, i.e. $d_{a_g}>1$ for all $a_g$. In this case, sometimes it is said that the symmetry defects carry non-Abelian zero modes (even when the topological order itself is Abelian).

    Without proofs we list three useful properties for quantum dimensions of symmetry defects:
    \begin{itemize}
    	\item First, define the total dimension of $g$ defects as
    \begin{equation}
    	\mathcal{D}_g=\sqrt{\sum_{a_g\in \mathcal{C}_g}d_{a_g}^2}.
    \end{equation}
    Then one can prove that $\mathcal{D}_g=\mathcal{D}$.
    \item Second, the number of $g$ defect types is the same as the number of $g$-invariant anyons, i.e. those $a$'s that obey $a=\varphi(a)$.
    \item Lastly, if all anyons are Abelian, then all $g$ defects must have the same quantum dimensions.

    \end{itemize}

	Intuitively, a defect loop can be thought of as the trajectory of a symmetry defect. That is, if one first creates a pair of defects $g$ and $g^{-1}$, then moves the $g$ defect along the defect loop, all the way until it is annihilated together with the $g^{-1}$ defect. 
	{
	An important subtlety is that in general we do not know what is the type of the defect loop created by $U_M(g)$. Most generally, the type of the defect can be represented as a ``superposition'' (direct sum to be more precise) of the ``irreducible" types:
	\begin{equation}
		\bigoplus_{a_g\in \cal{C}_g}n_{a_g} a_g.
		\label{gtype}
	\end{equation}
	Here $n_{a_g}$ are non-negative integers, called the multiplicity of the $a_g$ type. These integers, while quantized, are not completely universal. Namely, they are not uniquely fixed by the underlying symmetry-enriched phase of matter. We will determine them for a large class of systems from microscopic considerations in Sec. \ref{Sec:III}.
	}

	In the diagrammatic formalism for anyons (see Refs. [\onlinecite{Kitaev2006}] and [\onlinecite{SET}] for an introduction), 
	{
	the process of creating a $g$ defect loop of the type Eq. \eqref{gtype}
	}
	 is associated with an amplitude  
	{
	\begin{equation}
		d_g\equiv \sum_{a_g\in \cal{C}_g}n_{a_g}d_{a_g},
		\label{gdim}
	\end{equation}
	which is also the quantum dimension of this defect.
	}

	 It is then natural to postulate that the expectation value of $U_M(g)$, which also creates the same defect loop with the given defect type, is given by $d_{a_g}$ up to non-universal scaling factors. More precisely,
    \begin{equation}
    	|\langle U_M (g)\rangle| \approx d_{g} e^{-\alpha |\partial M|},
		\label{eqn:Ug1}
    \end{equation}
for a large, simply-connected region $M$, which naturally leads to Eq.~\eqref{eqn:tdp}. Here $\alpha$ is a non-universal constant.
We will show below that this is indeed the case in a broad class of examples.  

We illustrate the general theory outlined in this section with an example, which also makes connection with R\'enyi entropy. More examples will be given in Sec. \ref{Sec:IV}.

	Suppose the topological phase consists of $n$ identical layers, each of which is described by an anyon theory $\mathcal{C}$.  The anyon theory of the $n$ layers is denoted by $\mathcal{C}^{\boxtimes n}$.  Anyons are labeled by $n$-tuples $(a_1,a_2,\dots, a_n)$ where $a_i\in \mathcal{C}$. Since all the layers are identical, the system is invariant under any permutation of the layers, hence the symmetry group is the group of permutations $S_n$. Denote by $R$ the cyclic permutation:
	\begin{equation}
		R: (a_1, a_2,\dots, a_n)\rightarrow (a_2, a_3,\dots, a_1).
		\label{}
	\end{equation}
	The theory of $R$ defects has been well-understood, which we briefly review.

	Among all $R$ defects, there exists a``bare" defect, $0_R$, that satisfies the following fusion rule~\cite{XuCMP2004, Bischoff2019}:
\begin{equation}
	0_R\times \ol{0}_R=\sum_{a_1,\dots, a_n\in\mathcal{C}} N_{a_1a_2\cdots a_n}^0 (a_1,a_2,\dots, a_n).
\end{equation}
Here $N_{a_1a_2\cdots a_n}^0$ is the multiplicity of the vacuum $0$ in the tensor product $a_1\times a_2\times\cdots \times a_n$. This fusion rule can be understood intuitively as follows: in the presence of a $0_R-\ol{0}_R$ defect line, an anyon $(a_1,a_2,\dots, a_n)$ can be transformed into $(a_1\times a_2\times\cdots \times a_n,0,\dots,0)$ by moving the anyons around the $0_R$ defect and permute all of them to the same layer. If $ N_{a_1a_2\cdots a_n}^0>0$, then it means $(a_1,a_2,\dots, a_n)$ can be created out of the vacuum in the presence of a $0_R$ defect, which then implies the fusion rule.

Other defects can be obtained by
\begin{equation}
    a_R=(a,0,\cdots,0)\times 0_R,
\end{equation}
whose quantum dimension is $d_{a_R}=d_{0_R}d_a$. Then we find
\begin{equation}
	\cal{D}_R^2=\sum_{a\in \mathcal{C}}d_{a_R}^2 = d_{0_R}^2\sum_{a\in \mathcal{C}}d_a^2= d_{0_R}^2\cal{D}^2 .
\end{equation}
Since $\cal{D}_R=\cal{D}^{n}$, we must have $d_{0_R}=\cal{D}^{n-1}$. One can also directly evaluate $d_{0_R}$ from the fusion rule, the details of which can be found in Appendix.~\ref{app:genons}.
Defects of such layer permutation symmetry have been studied in the context of quantum Hall systems, known as ``genons"~\cite{BarkeshliPRX2012,BarkeshliPRB2013}.

Theoretically, the significance of cyclic permutation defects lies in the connection with R\'enyi entanglement entropy. It is well-known that the R\'enyi entanglement entropy can be computed using a replica trick~\cite{Calabrese_2004,Casini2006,Casini2011,Zhao2021,JRZhao2022,JRZhao2022QiuKu}. That is, for a quantum state $\ket{\psi}$, to compute the $n$-th R\'enyi entropy of a region $M$, one creates $n$ identical copies of the system, and define $R_M$ to be the cyclic permutation operator 
within $M$  among the $n$ copies. Then
    \begin{equation}
    	S^{(n)}(M)=\frac{1}{1-n} \ln\langle R_M\rangle,
    \end{equation}
	where the expectation value is taken over the state $\ket{\psi}^{\otimes n}$. We recognize that the R\'enyi entropy is essentially the logarithm of the disorder parameter of the replica symmetry~\cite{Calabrese_2004, Casini2006, Zhao2021}. According to our general formula, for a topological phase we expect that
	\begin{equation}
		S^{(n)}(M)=\frac{\alpha}{n-1} |\partial M|- \frac{1}{n-1}\ln d_R,
	\end{equation}
where $d_R$ is the quantum dimension of a certain $R$-defect. 		

Now suppose that the disorder operator $R_M$ indeed corresponds to the bare defect, which is supported by the CFT calculations below, according to the proposed formula Eq. \eqref{eqn:Ug1} the topological R\'enyi entropy is given by
\begin{equation}
 \gamma=\frac{1}{n-1}\ln d_{0_R}=\ln \mathcal{D},
\end{equation}
a well-known result~\cite{FlammiaPRL2009}.

We will now proceed to calculate the disorder parameter. We will first consider topological phases with gapless CFT boundary (more precisely, entanglement spectrum), and establish Eq. \eqref{eqn:Ug1} (with important corrections). Then we analyze several microscopic models to demonstrate the validity of the result even when the assumption of having CFT boundary does not hold.

\section{Topological phase with CFT entanglement spectrum}
\label{Sec:III}
We now present a derivation of Eq. \eqref{eqn:tdp} for topological phases whose entanglement spectrum can be described by a (1+1)d CFT. More precisely, the reduced density matrix of the ground state on a disk-like region $D$ is given by
\begin{equation}
	\rho_D= \frac{e^{-H_E}}{\Tr e^{-H_E}},
\end{equation}
where $H_E$ is the entanglement (modular) Hamiltonian. It has been conjectured and widely believed that $H_E$ belongs to the same universality class as the boundary theory of the topological phase~\cite{KitaevPreskill2006, LiHaldane}.  In many cases, the lower part of the entanglement spectra can be exactly matched with the low-energy spectrum of a physical edge up to overall rescaling~\cite{LiHaldane}. Ref. [\onlinecite{QiESPRL}] established the validity of the correspondence for general chiral phases under certain assumptions. We will thus assume for the remaining of this section that the entanglement Hamiltonian describes a CFT at low energy, which takes the same form as the edge one up to an overall scale.

For our derivation, the bulk-boundary correspondence plays a crucial role. Thus we first review how it works for chiral topological phase.  We assume that the boundary theory is a rational CFT, with the following Hamiltonian
\begin{equation}
	H_\text{edge} = \frac{2\pi v}{l}\left(L_0-\frac{c}{24}\right).
\end{equation}
Here $c$ is the chiral central charge and $l$ is the perimeter of the boundary. The Hilbert space of the boundary theory decomposes into a direct sum of superselection sectors $\mathcal{H}_a$, labeled by chiral primaries $a$. They are in one-to-one correspondence with anyon types in the bulk. When the system is a disk with no excitations in the bulk, the boundary CFT must be in the vacuum sector $\mathcal{H}_0$. To allow other superselection sectors, e.g. $\mathcal{H}_a$ on the boundary, there must be anyonic excitations whose total charge has type $a$ in the bulk.

For each chiral primary $a$, we define the character~\cite{YellowBook}
\begin{equation}
	\chi_a(\tau)=\Tr_{\mathcal{H}_a}e^{2\pi i\tau(L_0-\frac{c}{24})}.
	\label{}
\end{equation}
$\chi_a$ is essentially the Euclidean partition function over the superselection sector $\mathcal{H}_a$. Notably, there is not a single modular-invariant partition function of the theory. The characters $\chi_a$ transform under the modular transformations as
\begin{equation}
	\begin{split}
		\chi_a(\tau)&=\sum_bS_{ab}\chi_b(-1/\tau),\\
	\chi_a(\tau)&=\sum_b T_{ab}\chi_b(\tau+1)	
	\end{split}
	\label{eqn:charST}
\end{equation}
where $S$ and $T$ are the S and T matrices of the bulk anyon theory.

We now generalize this discussion to boundary theories that are not necessarily fully chiral~\cite{JiPRR2019, Ji2021}. Again we assume that the boundary is described by a CFT, which could be chiral or non-chiral. The Hamiltonian of the boundary theory is $H_\text{edge}=\frac{2\pi v}{l}H$, where 
\begin{equation}
	H=L_0+\ol{L}_0-\frac{c+\ol{c}}{24}
	\label{}
\end{equation}
is the dimensionless Hamiltonian of the CFT. Note that here we do not need to assume the left and right moving fields have the same chiral algebra. For example, the fully chiral case corresponds to $\ol{L}_0=0, \ol{c}=0$. We also define the momentum $P=L_0-\ol{L}_0$. The Hilbert space of the boundary theory splits into multiple superselection sectors, labeled by anyon types in the bulk~\cite{JiPRR2019}. For a superselection sector labeled by $a$, we can define the Euclidean partition function $\mathcal{Z}_a(\tau)$:
\begin{equation}
	\mathcal{Z}_a(\tau)=\Tr_{\mathcal{H}_a} e^{2\pi (i\tau_1 P-\tau_2 H)},
	\label{eqn:Za}
\end{equation}
where $\tau=\tau_1+i\tau_2$.
We remark that in general, there are no direct correspondence between the superselection sectors and the primary fields in the CFT. The partition functions $\mathcal{Z}_a$ also satisfy the relations Eq. \eqref{eqn:charST} under modular transformations.

Based on the assumptions laid out in the beginning of the section, we postulate that the entanglement Hamiltonian is given by
\begin{equation}
	H_E=\frac{\xi}{l}H,
	\label{}
\end{equation}
where $\xi$ is the correlation length of the bulk, and $l$ is the length of the disk. $H$ is again the dimensionless Hamiltonian of the CFT.
Note that if the disk $D$ does not contain any nontrivial excitations, then we must keep only the states in the vacuum superselection sector in the CFT. We will also assume that the system is bosonic, so we do not have to worry about subtleties related to spin structure.

Now consider a symmetry transformation $g$ of the bulk.  In the reduced density operator, suppose the symmetry transformation is represented by a unitary $U_g$. The disorder parameter can be evaluated as
\begin{equation}
	\langle U_g\rangle = \Tr U_g \rho_D=\frac{\Tr_{\mathcal{H}_0} U_g e^{-\frac{\xi}{l}H}}{\Tr_{\mathcal{H}_0} e^{-\frac{\xi}{l} H}}.
\end{equation}
Roughly speaking, the numerator is the CFT partition function with symmetry defect line inserted in the time direction. By a modular transformation, it can be related to the partition function over the $g$-twisted Hilbert space. It is therefore necessary to study the CFT in the defect sector. We denote by $\mathcal{H}_g$ the CFT Hilbert space on a spatial circle twisted by the $g$ symmetry. In general, $\mathcal{H}_g$ may also split into multiple superselection sectors, denoted by $\mathcal{H}_{a_g}$, where $a_g$ are precisely the defect types. We then define partition function with both spatial and temporal symmetry twists:
\begin{equation}
	\mathcal{Z}_{a_g}^{(g,h)}(\tau)=\Tr_{\mathcal{H}_{a_g}} U_he^{2\pi (i\tau_1 P-\tau_2 H)}.
	\label{}
\end{equation}
Namely, the partition function is the trace over the $\mathcal{H}_{a_g}$ defect sector and with $U_h$ inserted. Note that $\mathcal{Z}_a^{(\id,\id)}$ is nothing but $\mathcal{Z}_a$ defined in Eq. \eqref{eqn:Za}. 
{
Here and in the following we use $\id$ for the identity element of the group.  
}
In terms of these twisted partition functions, the disorder parameter is given by
{
\begin{equation}
	\langle U_g\rangle=\frac{\mathcal{Z}_{0}^{(\id,g)}(\frac{i\xi}{l})}{\mathcal{Z}_0^{(\id,\id)}(\frac{i\xi}{l})}.
	\label{eqn:Ug_Z}
\end{equation}
}

Similar to the $\mathcal{Z}_a$'s, the transformation properties of $\mathcal{Z}_{a_g}^{(g,h)}$ under modular group are entirely determined by the bulk. For our purpose, we need the following special case of S transformation:
{
\begin{equation}
	\mathcal{Z}_0^{(\id,g)}(\tau)=\sum_{a_g\in \mathcal{C}_g} \mathcal{S}_{\id, a_g}^{(\id,g)} \mathcal{Z}^{({g},\id)}_{a_g}(-1/\tau).
	\label{}
\end{equation}
}
Here $\mathcal{S}_{a_g,b_h}^{(g,h)}$ is the extended S transformation between $(g,h)$ and $(h,g^{-1})$ defect sectors of the (2+1)d topological phase on a torus~\cite{SET}. Notice that 
{
$\mathcal{S}^{(\id,\id)}$ 
}
reduces to the S matrix of the bulk anyons: 
{
$\mathcal{S}^{(\id,\id)}_{ab}=S_{ab}$.
}
According to [\onlinecite{SET}], we have
\begin{equation}
	{\mathcal{S}_{\id a_g}^{(\id,g)}}=\frac{d_{a_g}}{\mathcal{D}}.
	\label{}
\end{equation}

Since $\tau=\frac{i\xi}{l}$ and we are interested in Eq. \eqref{eqn:Ug_Z} in the limit $\xi\ll l$ (i.e. the high temperature limit for the reduced density operator), then $-1/\tau=\frac{il}{\xi}$ is effectively in the low temperature limit. Therefore we can expand ${\mathcal{Z}^{(g,\id)}_{a_g}(\frac{il}{\xi})}$ as a series of $e^{2\pi i\tau}=e^{-\frac{2\pi l}{\xi}}$:
\begin{equation}
	\begin{split}
		{\mathcal{Z}^{({g},\id)}_{a_g}\Big(\frac{il}{\xi}\Big)}&=\sum_{m=0}^\infty\sum_h p_h(m) e^{-\frac{2\pi l}{\xi}(h+m-\frac{c+\bar{c}}{24})}\\
		&\approx \sum_hp_h(0)e^{-\frac{2\pi l}{\xi}(h-\frac{c+\bar{c}}{24})}\\
		&\approx p_{h_{a_g}}(0) e^{-\frac{2\pi l}{\xi}(h_{a_g}-\frac{c+\bar{c}}{24})}.
	\end{split}
	\label{}
\end{equation}
Here $\sum_h$ means summing over primary fields in the in the defect sector $\mathcal{H}_{a_g}$ with conformal dimension $h$~\footnote{$h$ is the eigenvalue of $L_0+\ol{L}_0$.}, and $p_h(m)$ is the degeneracy of the level $h+m$. The degeneracy may come from different primaries having the same $h$, or a certain primary $h$ being a multiplet.
In the last step we only keep the one with the lowest conformal dimension, i.e. the highest weight state, which is denoted by $h_{a_g}$. We have thus found
\begin{equation}
	{\mathcal{Z}_0^{(\id,g)}\Big(\frac{i\xi}{l}\Big)} \approx\sum_{a_g\in \mathcal{C}_g} \frac{d_{a_g}}{\mathcal{D}} p_{a_g} e^{-\frac{2\pi l}{\xi}(h_{a_g}-\frac{c+\bar{c}}{24})}.
	\label{}
\end{equation}
Define $\Lambda_g$ as the set of defect sectors $a_g$ with the minimal $h_{a_g}$, among the entire $\mathcal{C}_g$, the corresponding value of the conformal dimension will be denoted by $h_g$. Define
\begin{equation}
	d_g= \sum_{a_g\in \Lambda_g}d_{a_g}p_{a_g}.
	\label{dg}
\end{equation}
Compared to Eq. \eqref{gdim}, we find that the multiplicity 
\begin{equation}
	n_{a_g}=
	\begin{cases}
		p_{a_g} & a_g\in \Lambda_g\\
		0 & \text{otherwise}
	\end{cases}.
\end{equation}

We can similarly evaluate the denominator:
\begin{equation}
	\begin{split}
	\mathcal{Z}_0\left( \frac{i\xi}{l} \right)&=\sum_{a\in \mathcal{C}}\frac{d_a}{\mathcal{D}}\mathcal{Z}_a\left( \frac{il}{\xi} \right)\\
	&\approx \sum_{a\in \mathcal{C}}\frac{d_a}{\mathcal{D}}p_a(0)e^{-\frac{2\pi l}{\xi}(h_a-\frac{c+\bar{c}}{24})}.
	\end{split}
	\label{}
\end{equation}
In the untwisted sector, the unique vacuum state with $h=0$ dominates the sum, so we obtain $\mathcal{Z}_0(\frac{i\xi}{l})\approx \frac{1}{\mathcal{D}}$. 
Putting the results together, the disorder parameter is given by
{
\begin{equation}
	\langle U_g\rangle=\frac{\mathcal{Z}_{0}^{(\id,g)}(\frac{i\xi}{l})}{\mathcal{Z}_0^{(\id,\id)}(\frac{i\xi}{l})}\approx d_g e^{-\frac{2\pi h_g}{\xi}l},
	\label{}
\end{equation}
}
plus exponentially small corrections. We thus find the topological disorder parameter is $\gamma_g=\ln d_g$. Compared with the proposed formula Eq. \eqref{eqn:Ug1}, $d_g$ accounts for the possibility that multiple defects could be ``degenerate''.

So far we have focused on the case of a disk-like region in the ground state. Practically it is often necessary to study systems on a cylinder or a torus, and the region may not be simply connected. We generalize the result to these situations in Appendix \ref{sec:TDP_on_cylinder}.

We conclude this section with the example of topological R\'enyi entropy discussed near the end of Sec.~\ref{Sec:SET}. Suppose that the topological phase $\mathcal{C}$ is fully chiral, thus having a chiral CFT boundary. The cyclic permutation orbifold of a chiral CFT has been studied in mathematical literature (e.g. [\onlinecite{XuCMP2004}]), and the conformal dimension of the highest weight state in the $R$-twisted sector $\mathcal{H}_{a_R}$ is given by $h_{a_R}=\frac{h_a}{n}+(n-\frac{1}{n})\frac{c}{24}$. Therefore the bare defect $h_{0_R}$ indeed has the lowest conformal dimension, without any additional degeneracy.

\section{ Examples}
\label{Sec:IV}

In this section we present detailed analysis of several examples. The motivation is two-folded: to demonstrate the result in concrete examples, and perhaps more importantly,
to study TDP when the boundary theory (or the entanglement spectrum) is not a CFT. This is particularly relevant for non-chiral topological phases, as the low-energy dynamics can be significantly different from a gapless CFT.

We will first investigate TDP in $\Z_N$ toric code models (Sec.~\ref{sec:IVA}) and quantum double models (Sec.~\ref{sec:IVB}), which represent important examples of non-chiral topological phases. In the exactly solvable limit, we compute analytically the TDP and show that the results agree with the general formula Eq.~\eqref{eqn:tdp}. We then employed DMRG simulations on finite cylinders to numerically study the models under external magnetic fields, which are no longer exactly solvable, and still find consistent results. Lastly, we study interesting examples of the TDP in Spin$(\nu)_1$ topological phases with microscopic O$(\nu)$ symmetry (Sec.~\ref{sec:IVD}) and the TDP for translation symmetry in Wen's plaquette model (Sec.~\ref{sec:IVC}) both analytically and numerically, where translation acts as electro-magnetic duality.

\subsection{Charge-conjugation symmetry in $\Z_N$ toric code}
\label{sec:IVA}

\begin{figure}[!t]
\includegraphics[angle=0,width=\linewidth]{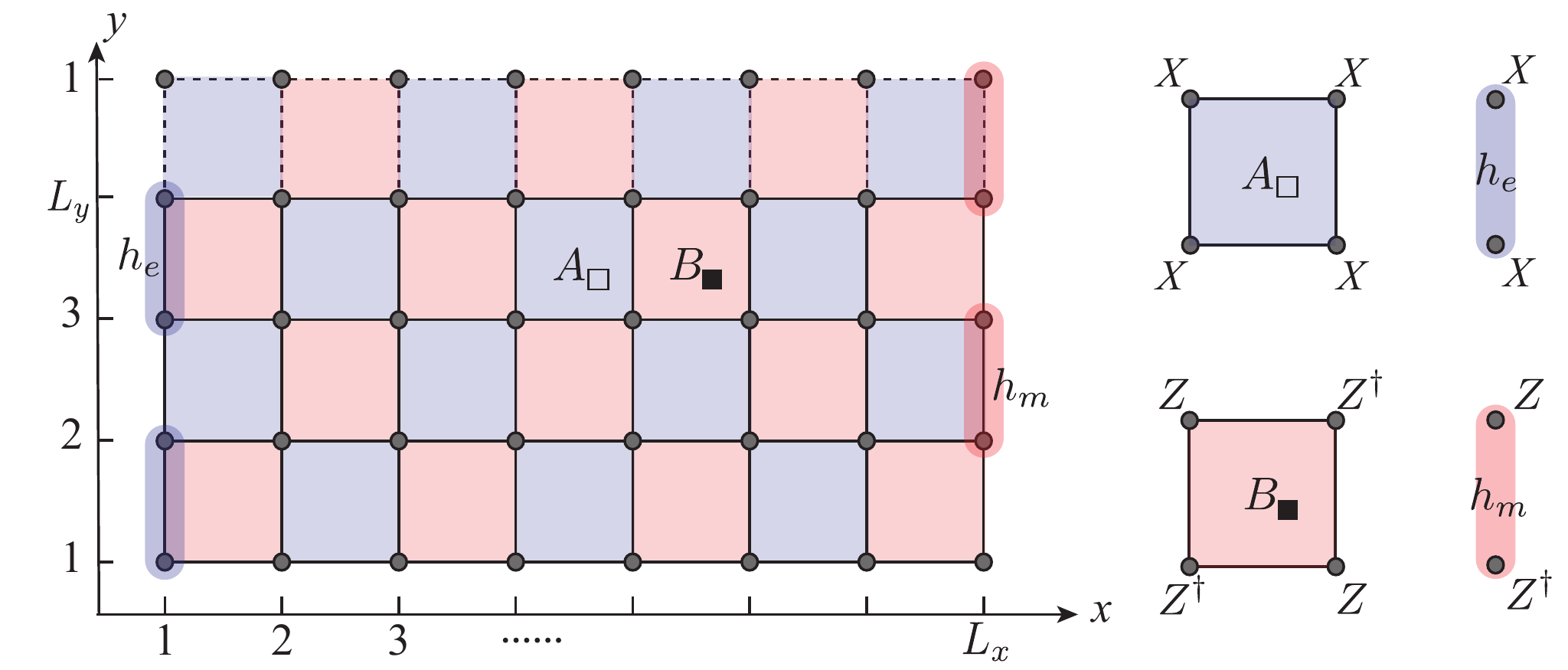}
\caption{The $\Z_N$ toric code model with the plaquette terms $A_\square = X_1X_2X_3X_4$,
$B_\blacksquare=Z^{\,}_1 Z^\dagger_2 Z^{\,}_3 Z^\dagger_4$, and the edge terms
$h_e = X^{\,}_{2j-1}X^{\,}_{2j}$, $h_m = Z^\dagger_{2j} Z^{\,}_{2j+1}$.
Here every site is labeled by its column index $x$ and row index $y$ with periodic condition $y+L_y\sim y$.}
\label{Fig:ZnToric}
\end{figure}

In this section we compute TDP for charge-conjugation symmetry in a $\Z_N$ toric code model. The model is defined on a checkerboard lattice in Fig.~\ref{Fig:ZnToric}, where each site has a $\Z_N$ spin. For one site, given an orthonormal basis $\ket{n}, n=0,1,\cdots, N-1$, we define the clock and shift operators:
\begin{equation}
	Z\ket{n}=\omega^n\ket{n}, X\ket{n}=\ket{[n+1]_N},
	\label{}
\end{equation}
where $\omega=e^{\frac{2\pi i}{N}}$ and $[\cdot]_N$ means $\cdot$ mod $N$. They obey the algebra
\begin{equation}
	\begin{gathered}
		Z_\mb{r}^N=X_\mb{r}^N=\mathds{1},
	Z_\mb{r}X_{\mb{r}}=\omega X_{\mb{r}}Z_{\mb{r}},\\
	\end{gathered}
	\label{}
\end{equation}
and commute on different sites.
The Hamiltonian in the bulk takes the following form:
\begin{equation}
	\begin{split}
		H=&-\sum_{\square}(A_\square+\text{h.c.})- \sum_{\blacksquare}(B_\blacksquare +\text{h.c.})\\
		&- \sum_{\mb{r}}(h_xX_\mb{r}+h_zZ_\mb{r}+\text{h.c.}),
\end{split}
	\label{eq:eq21}
\end{equation}
where $A_\square = X_1X_2X_3X_4$ and
$B_\blacksquare=Z^{\phantom{\dagger}}_1 Z^\dagger_2 Z^{\phantom{\dagger}}_3 Z^\dagger_4 $
as shown in \Fig{Fig:ZnToric}.
Notice that when $h_x=h_z=0$, the Hamiltonian consists of commuting terms and thus can be exactly solved. The ground state has all $A_\square=B_\blacksquare=1$ for all squares. There are two types of elementary excitations: an $e$ excitation corresponds to $A_\square=\omega$ for a certain $\square$, and a $m$ excitation corresponds to $B_\blacksquare=\omega$. All other excitations can be generated by forming bound states of multiple $e$'s and $m$'s. Since $A^N=B^N=1$, both $e$ and $m$ obey $\Z_N$ fusion rules. So there are altogether $N^2$ topologically distinct types of excitations, of the form $e^a m^b$, where $a,b\in \{0,1,\cdots, N-1\}$.

The Hamiltonian defined in Eq.~\eqref{eq:eq21} enjoys a charge-conjugation symmetry $U=\prod_\mb{r}U_\mb{r}$, which acts on the $\Z_N$ spin as
\begin{equation}
	C\ket{n}=\ket{N-n}.
	\label{Eq:UZ3}
\end{equation}
Notice that $U$ is the identity for $N=2$, so we will assume $N>2$ in the following.  It is easy to see that $C^2=\mathds{1}, CX_\mb{r}C^\dag = X_\mb{r}^\dag, CZ_\mb{r}C^\dag =Z_\mb{r}^\dag$, so $CA_\square C^\dag = A_\square^\dag = A_\square^{N-1}$, and similar for $B_\blacksquare$'s. Thus under the action of $U$, the excitations transform as
\begin{equation}
	C:e^a m^b \rightarrow e^{N-a}m^{N-b}.
	\label{}
\end{equation}
In other words, $U$ acts as charge conjugation.

Let us study the symmetry defects of $C$ for $N>2$. For odd $N$, there are no anyons that are invariant under $C$. Thus by the general results in Sec. \ref{Sec:SET}, there is only one type of symmetry defect, denoted by $\sigma_C$. Since the total quantum dimensions of the defects must be equal to that of the anyons, we find $d_{\sigma_C}=N$. One can show that $\sigma_C$ satisfies the following fusion rule:
\begin{equation}
	\sigma_C\times \sigma_C=\sum_{a,b=0}^{N-1}e^a m^b.
	\label{}
\end{equation}
For even $N$, there are four $C$-invariant anyons: $1, e^{N/2}, m^{N/2}, e^{N/2}m^{N/2}$, so there should be four distinct types of defects. Since the total dimension of defects must be $N$, one finds all defects have quantum dimension $\frac{N}{2}$.

To summarize, we find that the $C$ symmetry defects have quantum dimension
\begin{equation}
	d_{a_C}=\begin{cases}
     N & N\text{ odd}\\
    \frac{N}{2} & N\text{ even}
\end{cases}.
	\label{}
\end{equation}

We now compute the TDP in the model. First we will present analytical calculations in the exactly solvable limit $h_x=h_z=0$, and then use DMRG to study the model with the fields on.

\subsubsection{Exactly solvable limit}
\label{sec:Z3A}

We place the system on a finite cylinder of circumference $L_y$. Lattice sites are labeled by $x,y$ where $x=1,2,\cdots, L_x$ and $y=1,2,\cdots, L_y$, with periodic boundary condition $y\sim y+L_y$. As will be shown below, the Hamiltonian actually has degenerate edge states, which complicates the calculation. We can add additional terms on the boundary to lift these degeneracy, and there are actually two distinct choices for the boundary Hamiltonian
\begin{equation}
	\begin{split}
		H_\text{e}&=-\sum_{y \text{ odd}} (X_{y} X_{y+1}+\text{h.c.}),\\
		H_\text{m}&=-\sum_{y \text{ even}} (Z_{y}^\dag Z_{y+1}+\text{h.c.}).
	\end{split}
	\label{}
\end{equation}
Here we suppress the $x$ coordinates. See Fig. \ref{Fig:ZnToric} for illustrations of the boundary terms. Physically, $H_{e/m}$ condenses $e/m$ anyons on the boundary. In the following we turn on $H_e$ on the left boundary and $H_m$ on the right boundary, which has the additional effect of leaving only a unique ground state with a trivial anyon flux threading the cylinder.

First we determine the reduced density matrix corresponding to an entanglement cut along the $y$ direction. While the result is fairly well-known, we provide a derivation to be self-contained. Following Ref. [\onlinecite{HoPRB2015}], the Hamiltonian can be written as
\begin{equation}
	H=H_l+H_r+H_{lr},
	\label{}
\end{equation}
where $H_{l/r}$ are the Hamiltonian terms restricted entirely to the left and right halves of the cylinder, and $H_{lr}$ contains those defined on the plaquettes along the cut. Here we define $l$ to be all sites with $x\leq L_x/2$, and $r$ to be the half with $x>L_x/2$. We then characterize the ground-state subspace $\mathcal{V}_{l/r}$ of $H_{l/r}$, which can be defined from the algebra of observables commuting with $H_{l/r}$. In the following we consider the right half only. Such low-energy observables are necessarily localized on the two boundaries of the half cylinder. For the boundary at $x=L_x/2+1$, we find the algebra is generated by
\begin{equation}
	\begin{gathered}
	X_y X_{y+1}, \text{ for all {odd $y$'s} }\\
	Z_y Z_{y+1}^\dag, \text{ for all {even $y$'s} }.
	\end{gathered}
	\label{localalg}
\end{equation}
However, since we have fixed the cylinder ground state to be in the vacuum sector, the following two constraints must be imposed:
\begin{equation}
	X_1X_2\cdots X_L=Z_2Z_3^\dag Z_4Z_5^\dag\cdots Z_{L}Z_1^\dag = \mathds{1}.
	\label{Wilsonloop}
\end{equation}
Physically these two operators are the Wilson loops of $m$ and $e$ excitations wrapping around the cylinder, respectively.
We can think of the relations Eq. \eqref{Wilsonloop} as constraints that define the low-energy Hilbert space, which forbids operators like $X_y$ or $Z_y$, but the products defined in Eq. \eqref{localalg} commute with the constraints and are thus allowed operators in the theory.

Now we define new $\Z_N$ spin operators $\tilde{X}_j$ and $\tilde{Z}_j$ for $j=1,2,\cdots, L_y/2$:
\begin{equation}
	\tilde{X}_j \equiv X_{2j-1}X_{2j}, \tilde{Z}_j\tilde{Z}_{j+1}^\dag \equiv Z_{2j}Z_{2j+1}^\dag.
	\label{}
\end{equation}
These new spin operators satisfy
\begin{equation}
	\begin{gathered}
		\tilde{X}_j^N=\tilde{Z}_j^N=\mathds{1},\\
	[\tilde{X}_i, \tilde{X}_j]=[\tilde{Z}_i, \tilde{Z}_j]=0, \: i\neq j\\
	\tilde{X}_j \tilde{Z}_j=\omega \tilde{Z}_j\tilde{X}_j,\\
	[\tilde{X}_i, \tilde{Z}_j]=0, \: i\neq j
	\end{gathered}
	\label{}
\end{equation}
so they describe a chain of $\Z_N$ spins.  It is straightforward to verify that the constraint $Z_2Z_3^\dag Z_4Z_5^\dag\cdots Z_{L_y}Z_1^\dag = \mathds{1}$ is automatically satisfied, and the other constraint becomes:
\begin{equation}
	\prod_{j=1}^{L_y/2}\tilde{X}_j=\mathds{1}.
	\label{eqn:constraint}
\end{equation}
We thus conclude that the boundary Hilbert space is given by a $\Z_N$ spin chain subject to the global constraint Eq. \eqref{eqn:constraint}. The dimension of the Hilbert space is $N^{L_y/2-1}$. A similar analysis can be carried out for the left half of the cylinder. In the following we will denote by $\tilde{X}_{\eta j}$ and $\tilde{Z}_{\eta j}, \eta=l/r$ for the observables corresponding to the left/right halves of the cylinder.

Now we couple the left and right halves by $H_{lr}$, which after projection to the ground state subspace of $H_l$ and $H_r$ becomes
\begin{equation}
	H_{lr}=-\sum_{j=1}^{L_y/2}(\tilde{X}_{lj}\tilde{X}_{rj}+\tilde{Z}_{lj}\tilde{Z}_{l,j+1}^\dag\tilde{Z}_{rj}^\dag\tilde{Z}_{r,j+1}+\text{h.c.}).
	\label{}
\end{equation}
Thus the ground state must have $\tilde{X}_{lj}=\tilde{X}_{rj}^\dag, \tilde{Z}_{lj}\tilde{Z}_{l,j+1}^\dag =\tilde{Z}_{rj}\tilde{Z}_{r,j+1}^\dag$. In the eigenbasis of $\tilde{X}_{lj}$ and $\tilde{X}_{rj}$, one can show that the (normalized) wavefunction is given by
\begin{equation}
	\frac{1}{N^{\frac{L_y-2}{4}}}\sum_{\{\tau_j\}}\ket{\{\tau_j\}}_l\otimes \ket{\{\tau_j^*\}}_r,
	\label{}
\end{equation}
where $\tau_j$ is the eigenvalue of $\tilde{X}_j$ and the sum is restricted to those configurations with $\prod_{j=1}^{L_y/2}\tau_j=1$.

Tracing out half of the cylinder, say the left half, one finds that
\begin{equation}
	\rho_l=\frac{1}{N^{L_y/2-1}}\mathds{1}.
	\label{}
\end{equation}
Thus we reproduce a well-known result, that the reduced density operator in a stabilizer model describes a maximally mixed state with a completely flat entanglement spectrum.

Now we turn to the disorder operator $C_l$, which is $C$ restricted to the left cylinder. It is clear that $C_l$ is projected to the charge conjugation on the boundary Hilbert space:
\begin{equation}
	C_l\tilde{X}_{lj}C_l^\dag=\tilde{X}_{lj}^\dag, C_l\tilde{Z}_{lj}C_l^\dag=\tilde{Z}_{lj}^\dag.
	\label{}
\end{equation}
In the $\tilde{X}_{lj}$ eigenbasis, we have $C_l\ket{\{\tau_j\}}=\ket{\{\tau_j^*\}}$. The average of $C_l$ is given by
\begin{equation}
	\Tr (C_l\rho_l)=\frac{1}{N^{L_y/2-1}}\Tr C_l.
	\label{}
\end{equation}
At this point we need to distinguish odd and even $N$. For $N$ odd, only the state with $\tau_j=1$ is invariant under $C_l$, so $\Tr C_l=1$. Thus we find
\begin{equation}
    \Tr (C_l\rho_l) = \frac{1}{N^{L_y/2-1}}.
\end{equation}
In other words
\begin{equation}
	-\ln \langle C_l\rangle = \frac{\ln N}{2}L_y - \ln N,
	\label{Eq:UZ3Scale}
\end{equation}
which gives $\gamma=\ln N$.

For even $N$, there are $2^{L_y/2-1}$ basis invariant under $C_l$ where $\tau_j=\pm 1$, again subject to the constraint $\prod_j \tau_j=1$. Therefore
\begin{equation}
    \Tr (C_l\rho_l) = \frac{1}{(N/2)^{L_y/2-1}}.
\end{equation}
And the TDP is $\gamma=\ln \frac{N}{2}$.

We find that in both cases, $e^\gamma$ agrees with the quantum dimension of a single charge conjugation defect.

So far we have considered the exactly solvable limit. Once the external fields are turned on, the model is no longer exactly solvable. However, by adiabatic continuity, we expect that the boundary Hilbert space, defined by the algebra of ``low-energy'' observables, should remain the same. On the other hand, the density operator in general becomes a Gibbs state of a local entanglement Hamiltonian:
\begin{equation}
	\rho_l\propto e^{-H_E},
	\label{}
\end{equation}
subject to the global constraint Eq. \eqref{eqn:constraint}. In other words, $H_E$ is a local Hamiltonian that commutes with the global constraint \eqref{eqn:constraint}. Below we consider two examples.

First we assume $H_E=\beta\sum_j (\tilde{X}_j+\text{h.c.})$, which may be a reasonable approximation for the model with a small $h_x$ and $h_z=0$. First we compute the partition function $\mathcal{Z}_l=\Tr P_0e^{-H_E}$, where $P_0$ is the projector to the space with $\prod_j \tilde{X}_j=1$. We write the projector as
\begin{equation}
    P_0=\frac{1}{N}\sum_{k=0}^{N-1}\prod_j \tilde{X}_j^k.
\end{equation}
 The partition function is
\begin{equation}
	\begin{split}
	\mathcal{Z}_l&=\frac{1}{N}\sum_{k=0}^{N-1} \prod_j\Tr(\tilde{X}_j^k e^{-\beta (\tilde{X}_j+\text{h.c.})})\\
	&=\frac{1}{N}\sum_{k=0}^{N-1}\left[\sum_{p=0}^{N-1}e^{\frac{2\pi ipk}{N}} e^{-2\beta\cos \frac{2\pi p}{N} }\right]^{L_y}
	\end{split}
\end{equation}
Then we can evaluate
\begin{equation}
	\Tr (P_0C e^{-H_E}) = e^{-2\beta L_y}.
\end{equation}
We numerically evaluate $\langle C\rangle$ for various values of $\beta$, and in all cases obtain the same TDP as the $\beta=0$.

As a second example, we consider what happens when both $h_x$ and $h_z$ magnetic fields are turned on. In this case the dynamics becomes more complicated. One possibility is that the entanglement Hamiltonian is tuned to a critical point described by a $\Z_3$ parafermion CFT. More concretely, suppose that the entanglement Hamiltonian can take the following form:
\begin{equation}
	H_E= -\beta\sum_j (\tilde{Z}_j^\dag \tilde{Z}_{j+1}+\tilde{X}_j+\text{h.c.}), \beta>0,
	\label{}
\end{equation}
which is the critical point of the $\Z_3$ Potts model. Suppose that we restrict ourselves to the CFT Hilbert space. We will write the partition functions in terms of the characters $\chi_h$ of the minimal model $\mathcal{M}(6,5)$, where $h$ labels the conformal dimension of the primary. In fact, $\mathcal{M}(6,5)$ can be obtained from the $\Z_3$ parafermion CFT by orbifolding the $\Z_2$ charge conjugation symmetry. The partition function of the boundary CFT in the vacuum sector is given by~\cite{Ji2021}
\begin{equation}
	\mathcal{Z}_{0}=|\chi_0+\chi_3|^2+|\chi_{\frac25}+\chi_{\frac75}|^2.
	\label{}
\end{equation}
Note that it is different from the modular-invariant partition function of a genuine (1+1)d $\Z_3$ parafermion CFT.
$\chi_3$ can be understood as the $\Z_2$ charge sector. Then the partition function with $\Z_2$ symmetry operator inserted in the time direction can be easily written down:
\begin{equation}
	\mathcal{Z}_{0}^{(\id,C)}=|\chi_0-\chi_3|^2+|\chi_{\frac25}-\chi_{\frac75}|^2.
	\label{}
\end{equation}
Ref. [\onlinecite{Ji2021}] showed that there is a single defect sector with the following partition function
\begin{equation}
	\begin{split}
		\mathcal{Z}_{\sigma_C}^{(C,\id)}&=|\chi_{\frac18}+\chi_{\frac{13}{8}}|^2+|\chi_{\frac{1}{40}}+\chi_{\frac{21}{40}}|^2.\\
\end{split}
	\label{}
\end{equation}
So the field with smallest conformal dimension is non-degenerate, with $h=(\frac{1}{40},\frac{1}{40})$. In other words, $p_{\sigma_C}=1$. Using Eq. \eqref{dg}, we conclude that $d_C=3$ in this case.

\subsubsection{DMRG results}
\label{sec:IVA2}

\begin{figure}[!t]
\includegraphics[angle=0,width=\linewidth]{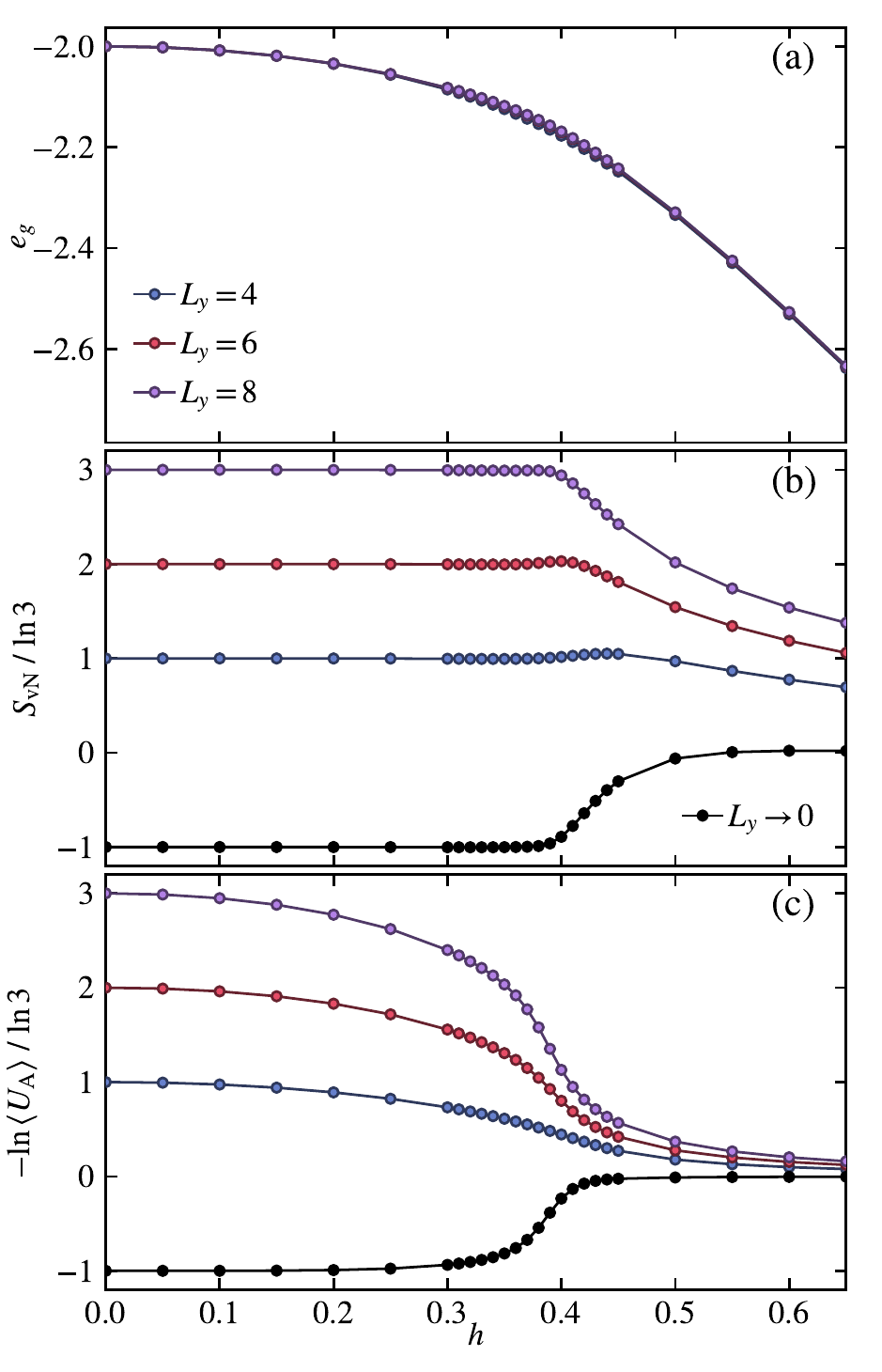}
\caption{DMRG results for the $\Z_3$ toric code model with cylinder geometry, (a) ground-state energy $e_g$, (b) entanglement entropy (EE) $S_{\mathrm{vN}}$, and (c) the negative logarithmic value of disorder operator $-\ln|{\langle U_{\mathrm{A}}\rangle}|$ of subsystem A are shown versus the fields $h_x=h_z=h$. The black dots denote the $L_y\rightarrow0$ extrapolated values
showing finite value $-\ln{3}$ in $\Z_3$ topological ordered phase. }
\label{Fig:Z3TDP}
\end{figure}

For systems away from exactly solvable limit and perturbative regime, we employ the DMRG algorithm to compute the TDP of the $\Z_3$ toric code with both transverse and longitudinal fields for the Hamiltonian in Eq.~\eqref{eq:eq21}. The cylinder geometry for the DMRG computations is illustrated in the left panel in Fig.~\ref{Fig:ZnToric}, and we have studied cylinders with $L_x=16$ and $L_y=4,6,8$. Boundary terms are introduced to lift the degeneracy as discussed in Sec.~\ref{sec:Z3A}, shown in Fig.~\ref{Fig:ZnToric} right panels. We have retained up to $D=512$ bond states in the simulations which ensure the discarded weight is at most $O(10^{-5})$.

In \Fig{Fig:Z3TDP} (a), we show the ground-state energy
density $e_g = \tfrac{1}{L_xL_y} \langle \psi_{\mathrm{gs}}| H |\psi_{\mathrm{gs}}\rangle $
versus the fields $h_x = h_z = h$, where $|\psi_\mathrm{gs}\rangle$ is the DMRG ground state.
The derivative of $e_g$ with respect to $h$ changes abruptly around $h_x=h_z=h_c \simeq 0.4$, indicating a first-order phase transition from the topologically ordered phase to the trivial phase
around $h_c$, in good agreement with the previous study \cite{Schulz2012}.

To verify the topological features of both phases,
we calculate the bipartite (von Neumann) entanglement entropy (EE),
$S_{\mathrm{vN}} = - \mathrm{tr}( \rho_A \ln{\rho_A} )$,
where $\rho_A =\mathrm{tr}_{\bar A}( |\psi_{\mathrm{gs}}\rangle\langle\psi_{\mathrm{gs}}| )$
is the reduced density matrix of subsystem $A$ and the bipartition is taken
such that the cylinder is cut vertically into two shorter cylinders
$A$ (the left half) and $\bar A$ (the right half) with length $L_x/2$ each.
As shown in \Fig{Fig:Z3TDP}(b), the EE data show clear area-law behaviour with a constant correction,
i.e., $S_{\mathrm{vN}} = \alpha L_y - \gamma$, with topological EE
$\gamma$ being $\ln{3}$ in small-$h$ topological phase and $0$ in large-$h$ trivial phase.

We now calculate the TDP for the charge conjugation symmetry. The disorder parameter is the ground-state expectation value of the corresponding disorder operator
$U_{\mathrm{A}} = \Pi_{\mathbf{r}\in \mathrm{A}} C_{\mathbf{r}}$ [c.f. \Eq{Eq:UZ3}], which should scale as $-\ln|{\langle U_{\mathrm{A}} \rangle}| = \alpha' L_y - \ln{3}$ [c.f. \Eq{Eq:UZ3Scale}] in the topologically ordered phase and approach zero in the trivial phase.
In \Fig{Fig:Z3TDP}(c), we explicitly show such behaviour in the small-$h$ phase. As $h$ increases, the non-universal coefficient $\alpha'$ deviates away from the $h=0$ value $\frac{\ln 3}{2}$, the constant correction remains robustly $-\ln 3$ throughout the topological phase. In the topologically trivial large-$h$ phase, DP indeed extrapolates to zero.

\subsection{Quantum double model}
\label{sec:IVB}
In this section we study the quantum double model of a finite group $G$, which can be regarded as a $G$ lattice gauge theory. The model has a $G$ spin on each edge of the lattice, which can be thought of as a regular representation of $G$. We will not give details of the Hamiltonian, which can be found in e.g. Ref. [\onlinecite{KitaevQD}]. It is essentially a microscopic realization of a $G$ gauge theory, where the $G$ spins play the role of $G$ gauge fields. When $G=\Z_N$, the model reduces to the $\Z_N$ toric code as discussed in the previous section. For a non-Abelian group $G$, the quantum double model describes a non-Abelian topological phase. The topological order of the quantum double model is denoted as D$(G)$, mathematically described by the Drinfeld center of the fusion category Vec$_G$ (or the Morita equivalent one Rep($G$))~\cite{BuerschaperPRB2009}. Physically, D$(G)$ describes anyons carrying electric and magnetic charges of the gauge group $G$.

We shall consider the following symmetry transformation:
\begin{equation}
    U\ket{g}=\ket{\varphi(g)},
\end{equation}
where $\varphi$ is an automorphism of the group $G$. It can be shown that $U$ is a symmetry of the quantum double model.
The symmetry action of $U$ on the anyons is naturally induced from the group automorphism $\varphi$.  For later use we define
\begin{equation}
    G_\varphi=\{g\in G|\varphi(g)=g\},
\end{equation}
and $n_\varphi=|G_\varphi|$.

We now proceed to calculate the disorder parameter for the $U$ symmetry. Following a procedure similar to the derivation in the $\Z_N$ case, one can show that the boundary Hilbert space is a one-dimensional $G$ spin chain, restricted to the $G$-invariant subspace~\cite{Albert2111}.  More specifically, each site of the spin chain forms a regular representation of $G$, with basis $\{|\{g_j\}\rangle\}_{ g_j\in G}$. For each $g\in G$, define a $g$ global symmetry in the spin chain as
\begin{equation}
	L_g\ket{\{g_j\}}=\ket{\{gg_j\}}.
	\label{}
\end{equation}
We then demand that $L_g\equiv \mathds{1}$ in the Hilbert space. It is useful to define a $G$-invariant basis as follows:
\begin{equation}
	\ket{g_1,g_2,\cdots,g_{L-1}}'\equiv\frac{1}{|G|}\sum_{h\in G}\ket{\{hg_1,  \cdots, hg_{L-1},h\}}.
	\label{}
\end{equation}
Here we only keep $g_1$ to $g_{L-1}$, as $g_L$ is redundant once summing over the entire $G$ orbit. Put it in another way, we pick a representative in the orbit with $g_L=1$. We will refer to $\ket{\{g_1,g_2,\cdots, g_{L-1}\}}$ as the $G$-invariant basis.

In the fixed-point ground state wavefunction, the reduced density matrix is the maximally mixed state
 $\rho =\frac{1}{|G|^{L-1}} \mathds{1}$.

The symmetry transformation $U$ naturally restricts to essentially the same transformation on the spin chain:
\begin{equation}
	U_M\ket{\{g_j\}}=\ket{\{\varphi(g_j)\}}.
	\label{}
\end{equation}
On the $G$-invariant basis we find
\begin{equation}
	\begin{split}
	U_M\ket{\{g_j\}}'&=\frac{1}{|G|}\sum_{g\in G}\ket{\{\varphi(g)\varphi(g_j)\}}\\
	&=\frac{1}{|G|}\sum_{g\in G}\ket{\{g\varphi(g_j)\}}\\
	&=\ket{\{\varphi(g_j)\}}'.
	\end{split}
	\label{}
\end{equation}
Tracing over $\rho$, only states invariant under $\varphi$ can contribute, which adds up to $n_\varphi^{L-1}$. Thus
\begin{equation}
    \langle U_M\rangle=\left(\frac{n_\varphi}{|G|}\right)^{L-1},
\end{equation}
and we find
\begin{equation}
    \gamma = \ln \frac{|G|}{n_\varphi}.
\end{equation}
Note that $|G|/n_\varphi$ is always an integer.

Now let us compute the quantum dimensions of the $U$ symmetry defects. To this end, suppose that $\varphi$ is an order $r$ element in Aut$(G)$. Then we view $U$ as the generator of a $\Z_r$ group, and consider gauging the $\Z_r$ symmetry. Because of the nontrivial action of $U$ on the gauge group $G$, we end up with a new gauge theory with a larger gauge group $G\rtimes_\varphi \Z_r$. In other words, the gauged topological order is identified as D$(G\rtimes_\varphi \Z_r)$. The $U$ defects are promoted to gauge fluxes once the global symmetry is gauged. Since an Abelian symmetry is gauged, the quantum dimension of the gauge flux is the same as that of the defect.

We now need to analyze the anyon content of D$(G\rtimes_\varphi \Z_r)$. Recall that for a general finite group $H$, anyon types in D$(H)$ are labeled by a pair $([h],\pi_h)$, where $[h]$ is a conjugacy class with $h$ being a representative element, and $\pi_h$ is an irreducible representation of ths centralizer group $C_h$. Physically, $[h]$ labels the gauge flux and $\pi_h$ is the gauge charge attached to the flux. The quantum dimension of this anyon is $|[h]|\cdot \mathrm{dim}\, \pi_h$, where $|[h]|$ is the size of the conjugacy class.

Label the group elements of $G\rtimes_\varphi \Z_r$ by $(g, a)$ where $g\in G$, $a\in \Z/r\Z$. Then the multiplication  in $g\rtimes_\varphi \Z_r$ becomes
\begin{equation}
   (g,a)\cdot (h,b)=(g \varphi^a(h), [a+b]_r).
\end{equation}
We have $ {(g,a)}^{-1}=(\varphi^{a}(g^{-1}),[-a]_r)$.
Thus
\begin{equation}
	(g,a)\cdot (1, 1)\cdot {(g,a)}^{-1}=(g\varphi(g)^{-1}, 1).
\end{equation}
The conjugacy class of $(1,1)$ is then the quotient of $G$ by the subgroup $G_\varphi$. We also need to attach representations to the conjugacy class $[(1,1)]$. Choose $(1,1)$ as the representative, the stabilizer group is isomorphic to $G_\varphi\times \Z_r$, whose representations can be obtained from those of $G_\varphi$ and of $\Z_r$. The $\Z_r$ representations are just the $\Z_r$ gauge charges.  We thus conclude that $U$ defects are in one-to-one correspondent with irreducible representations of $G_\varphi$. For an irreducible representation $\pi$, the corresponding symmetry defect has quantum dimension $\frac{|G|}{n_\varphi}\cdot \mathrm{dim}\,\pi$. In particular, if we choose $\pi=\mathds{1}$ the identity representation, the quantum dimension is $\frac{|G|}{n_\varphi}$, which is the minimum among all defects. The total dimension of the defects is
\begin{equation}
	\frac{|G|^2}{n_\varphi^2}\sum_{\pi\in\mathrm{Rep}(G_\varphi)}(\mathrm{dim}\,\pi)^2=|G|^2,
	\label{}
\end{equation}
as expected. Therefore we find that $e^\gamma$ is equal to the minimal quantum dimension of the $U$ defects in D$(G)$.

\subsection{Spin$(2n)_1$}
\label{sec:IVD}

In this section we study TDP of a $\Z_2$ anyon-permuting symmetry in Spin$(2n)_1$ topological phase. There are four types of Abelian anyons, denoted by $1, \psi, v_+$ and $v_-$ with topological twist factors $\theta_\psi=-1,\theta_{v_\pm}=e^{i\frac{\pi n}{4}}$. The theory can be obtained from coupling fermionic topological superconductors with Chern number $2n$ to a $\Z_2$ gauge field, where $v_\pm$ correspond to fermion parity fluxes~\cite{Kitaev2006}. Here we will take the $n=0$ theory to be the $\Z_2$ toric code.
It is easy to see that there is a $\Z_2$ symmetry that swaps the $v_\pm$ anyons. Physically, the symmetry can be realized as follows for $n>0$: a topological superconductor with $C=2n$ is equivalent to $2n$ identical copies of $p+ip$ superconductors, which has a SO($2n$) symmetry that rotates the layers. A fermion parity flux through this system binds $2n$ Majorana zero modes $\eta_k, k=1,2,\cdots, 2n$.
Fixing the local fermion parity $i^n \eta_1\eta_2\dots\eta_{2n}$, there is a $2^{n-1}$-dimensional zero-energy subspace, that forms a spinor representation of SO($2n$) group. The total fermion parity of the Majorana zero modes can be even or odd, corresponding to the two types of fluxes $v_\pm$. Thus swapping the two types of fluxes is equivalent to flipping the fermion parity of the flux,  which can be achieved with the symmetry transformation $(-1)^{N_1}$ where $N_1$ is the fermion number in the first layer. Under this transformation, $\eta_1\rightarrow -\eta_1$ while the other Majoranas do not transform, so the fermion parity changes sign. This additional $\Z_2$ symmetry generated by $(-1)^{N_1}$ combines with SO$(2n)$ to form O$(2n)$ group.
Recently, a family of exactly solvable generalizations of Kitaev's $\Z_2$ spin liquid was introduced~\cite{ChulliparambilPRB2020, ChulliparambilPRB2021}, that realizes all Spin$(\nu)_1$ topological phases for any integer $\nu\geq 1$, and notably the O$(\nu)$ symmetry is realized exactly in the lattice model.

For $n=0$, it is customary to rename $v_\pm$ as $e$ and $m$, which can be thought of as the (bosonic) electric charge and magnetic charge of a $\Z_2$ gauge theory. The symmetry $e\leftrightarrow m$ is often called the electromagnetic duality (EDM). Analogous to the construction for Spin$(2n)_1$ with $n>0$, the EDM can be realized as follows: consider two layers, one forms a $p+ip$ superconductor and the other $p-ip$. Together the total Chern number is 0, so coupling to a $\Z_2$ gauge field results in a $\Z_2$ toric code. Through an almost identical analysis, one can see that the symmetry $(-1)^{N_1}$ permutes the two types $e$ and $m$ of fermion parity fluxes. This construction also suggests the Ising CFT as a possible symmetry-preserving edge theory. In fact, the construction still works if we replace the $p+ip$ superconductor with one that has an odd Chern number $2r+1$ (and the other layer in the mirror image). In that case, the edge theory is the non-chiral Spin$(2r+1)_1$ CFT.

In all these cases, both $1$ and $\psi$ anyons are invariant under the permutation. Thus there are two types of symmetry defects $\sigma_\pm$, which satisfy the Ising-like fusion rules:
\begin{equation}
	\sigma_\pm\times\sigma_\pm = 1+\psi, \sigma_+\times v_\pm = \sigma_-.
	\label{}
\end{equation}
Their quantum dimensions are $d_{\sigma_\pm}=\sqrt{2}$.

We proceed to compute the disorder parameter for the $v_+\leftrightarrow v_-$ symmetry. For $n>0$, the boundary theory models are chiral Spin($\nu)_1$ Wess-Zumino-Witten CFTs, which can be equivalently described as $\nu$ chiral Majorana fermions coupled to a $\Z_2$ gauge field~\cite{WittenNABosonization, Kitaev2006}. We calculate the disorder parameter directly using the chiral CFT and find
\begin{equation}
	\gamma=
	\begin{cases}
		\ln 2\sqrt{2} & n=1\\
		\ln \sqrt{2} &  n>1
	\end{cases}.
	\label{}
\end{equation}
Details of the calculation can be found in Appendix \ref{app:CFTpartial}. To understand why $n=1$ is special,  notice that the CFT Spin$(2)_1$ is equivalent to U(1)$_4$, whose $\Z_2$ orbifold is two copies of Ising CFT. Thus there are two defect primaries with the same conformal dimension $1/16$, contributing the extra factor of $2$ according to the general formula. For $n>1$ no such degeneracy of operator spectrum is present.

We can perform a similar calculation for the $\Z_2$ toric code. When the edge theory is the (non-chiral) Spin$(2r+1)_1$, we show in Appendix \ref{app:CFTpartial} that the TDP
\begin{equation}
	\gamma=\ln 2^{r+1}\sqrt{2}.
	\label{}
\end{equation}
Note that the value of $\gamma$ is different for different edge theories. The factor $2^{r}$ arises because the defect carries a spinor representation of SO$(2r+1)$, which begs for the question that whether the result is robust against perturbations that break the exact SO$(2r+1)$ symmetry of the CFT. In Appendix \ref{app:CFTpartial} we introduce velocity anisotropy to the Spin$(2r+1)_1$ CFT (i.e. different modes have different velocities), and show that such anisotropy does not affect the value of $\gamma$. In fact, even with such anisotropy, the spinor degeneracy of the defect still remains, which explains the robustness of $\gamma$.

For illustration, let us consider the $r=0$ case, where the edge theory is the Ising CFT.  The character in the vacuum sector reads
\begin{equation}
	\mathcal{Z}_0=|\chi_1|^2+|\chi_\psi|^2.
	\label{}
\end{equation}
For the definitions of the Ising characters see Appendix \ref{app:CFTpartial}. The defect sector partition functions have been obtained in Ref. [\onlinecite{Ji2021}]. There are two defect types, which will be denoted as $\sigma_\pm$, with the following partition functions:
\begin{equation}
	\begin{split}
		\mathcal{Z}^{(1,0)}_{\sigma_+}=\chi_\sigma(\ol{\chi}_1+\ol{\chi}_\psi),\\
		\mathcal{Z}^{(1,0)}_{\sigma_-}=({\chi}_1+{\chi}_\psi)\ol{\chi}_\sigma.
	\end{split}
	\label{}
\end{equation}
So the lowest conformal dimension $1/16$ is two-fold degenerate, contributing an additional factor of $2$.  More details can be found in Appendix \ref{app:CFTpartial}.

\subsection{Partial translation in Wen's plaquette model}
\label{sec:IVC}
\begin{figure}[!t]
\includegraphics[angle=0,width=\linewidth]{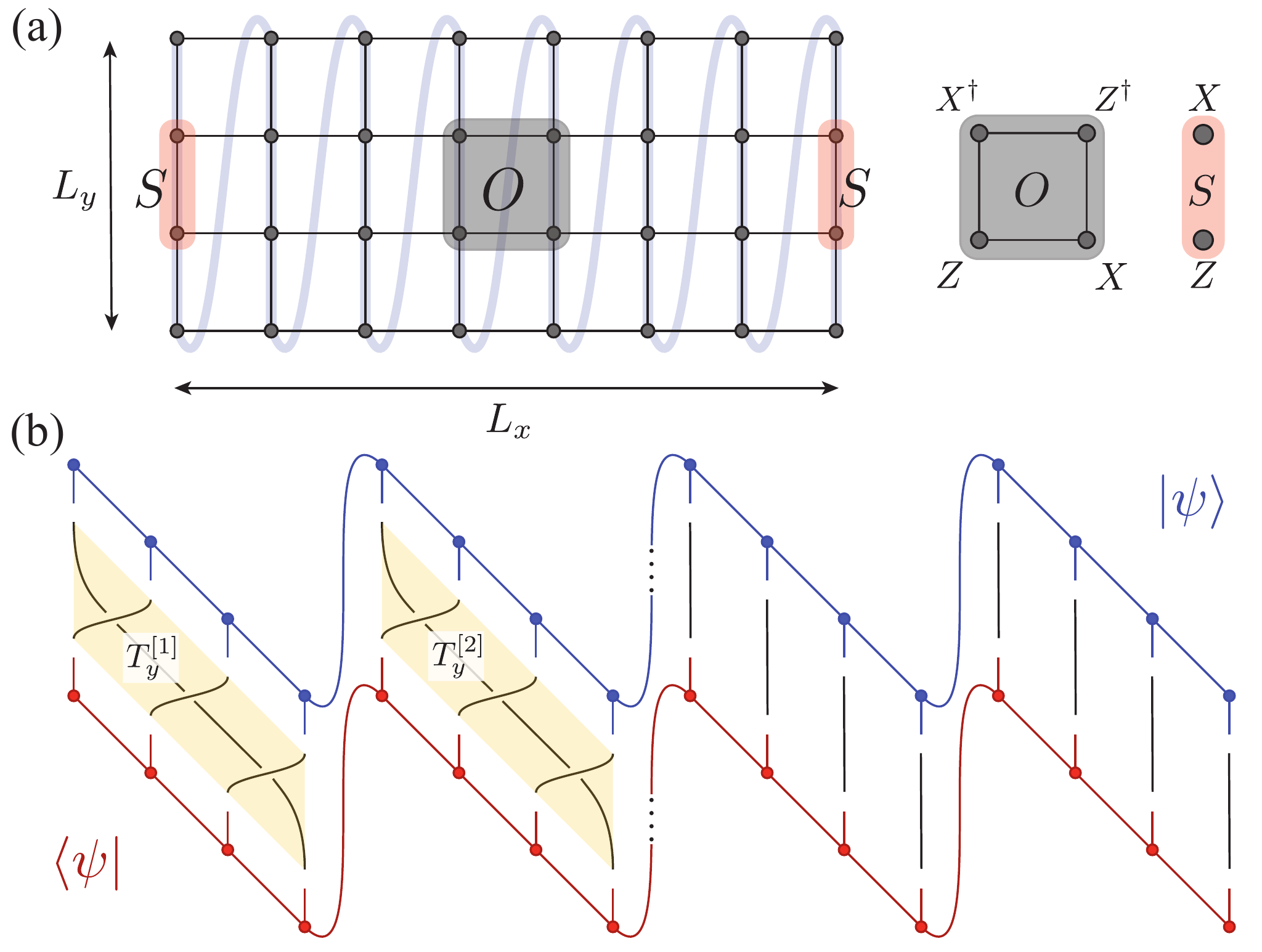}
\caption{(a) Wen's plaquette model with the plaquette term $O_\square = X^\dagger_1 Z^{\,}_2 X^{\,}_3 Z^\dagger_4 + \mathrm{H.c.}$, and
the edge term $S = XZ$. The blue-shaded line denotes the DMRG snake-like path in computation.
(b) Illustration of the partial translation operation $\bigotimes_{i=1}^{L_x/2} T_y^{[i]}$ as a disorder operator.  }
\label{Fig:WenPlaq}
\end{figure}

So far we have only considered on-site symmetries in our examples. We now demonstrate that the same idea can be applied to spatial symmetry.

The example we will consider is Wen's plaquette model \cite{WenPRL2003, You2012WenPlaquette}, which is a slight variation of the $\Z_N$ toric code model discussed in Sec.~\ref{sec:IVA}. Consider $\Z_N$ spins on a square lattice with the following Hamiltonian:
\begin{equation}
H=-\sum_{\square}  (O_\square + \text{H.c.}) - h \sum_i ( X_i +  Z_i + \mathrm{H.c.}),,
	\label{eq:eq61}
\end{equation}
where $ O_\square =  X^\dagger_1  Z^{\phantom{\dagger}}_2  X^{\phantom{\dagger}}_3  Z^\dagger_4 $ as denoted in \Fig{Fig:WenPlaq}(a).
We can see that the model is in fact unitarily equivalent to the one defined in {\Eq{eq:eq21}}, but now is completely invariant under translations on the square lattice. The $e$ and $m$ excitations are supported on the two types of plaquettes, thus transformed into each other by unit translations of the lattice. Since $e$ and $m$ can be viewed as electric and magnetic charges of an emergent $\Z_N$ gauge theory, such a symmetry is called the electromagnetic duality (EMD).

A lattice dislocation can be thought of as a defect for translation symmetry~\cite{BombinPRL2010}. It was shown in Ref. [\onlinecite{BombinPRL2010}] for $N=2$ and further in Ref. [\onlinecite{You2012WenPlaquette}] that dislocations in the plaquette model have quantum dimensions $\sqrt{N}$. There are $N$ topologically distinct types of dislocations, labeled by $\sigma_p$, which satisfy the fusion rule
\begin{equation}
	\sigma_0\times\sigma_0=\sum_{n=0}^{N-1}e^{-n}m^n, \sigma_0\times e^p = \sigma_0\times m^p=\sigma_p.
	\label{}
\end{equation}

In a planar geometry it is not clear how to apply translation to a finite region, since no finite region can be invariant under translation. Instead we consider the cylindrical geometry, where one can naturally apply translation along the periodic direction to only half of the cylinder. Below we will investigate the TDP associated with such partial translations in Wen's plaquette model. We first consider the $N=2$ case where the algebra is relatively simple and then generalize to other $N$.

\subsubsection{Exactly solvable limit}
\label{sec:IVC1}
We start from the exactly solvable point without any external fields. We use the same method as Sec.~\ref{sec:Z3A} to calculate the disorder parameter. In fact, the derivation of the boundary Hilbert space and the reduced density operator can be rather straightforwardly adopted here.  We find that the boundary Hilbert space is determined by the following observables:
\begin{equation}
	S_j=X_j Z_{j+1}, j=1,2,\cdots, L_y,
	\label{}
\end{equation}
with the following constraint imposed:
\begin{equation}
	\prod_{j=1}^{L_y/2}S_{2j}=\prod_{j=1}^{L_y/2}S_{2j-1}=\mathds{1}.
	\label{}
\end{equation}
Again we suppress the $x$ coordinate.

The $S_j$'s satisfy the algebra $S_j^2=\mathds{1}, S_jS_{j+1}=-S_{j+1}S_j$ (and otherwise commute). The translation symmetry acts in the obvious way:
\begin{equation}
T_yS_jT_y^{-1}=S_{j+1}.
	\label{}
\end{equation}

This boundary Hilbert space can be mapped to the $\Z_2$ symmetric sector of an Ising spin chain:
\begin{equation}
	S_{2j}\equiv \tau_j^x, S_{2j+1}\equiv \tau_j^z\tau_{j+1}^z,
	\label{}
\end{equation}
with the global constraint $\prod_{j=1}^{L_y/2}\tau_j^x=\mathds{1}$. However, now the translation symmetry acts as the Kramers-Wannier duality, which is difficult to handle in the spin representation.

To proceed, it is most convenient to ``fermionize'' this Hilbert space as a chain of Majorana operators. In the following we define $L\equiv L_y$. The fermionization map is basically the Jordan-Wigner transformation:
\begin{equation}
	\begin{gathered}
		S_j\equiv i\psi_j\psi_{j+1}, 1\leq j<L,\\
		S_{L}\equiv -i\psi_{L}\psi_1.
	\end{gathered}
	\label{}
\end{equation}
Here the Majorana operators satisfy $\{\psi_i,\psi_j\}=2\delta_{ij}$.
The global constraints after fermionization become
\begin{equation}
i^{L/2}\psi_1\psi_2\cdots\psi_{L_y}=1.
	\label{}
\end{equation}
The minus sign in $S_{L}$ is necessary in order to satisfy both constraints. The dimension of the Hilbert space is $D=2^{L/2-1}$. Following essentially the same steps as those in Sec. \ref{sec:Z3A}, one can show that the reduced density matrix is the maximally mixed state.

The translation symmetry $T_y$ acts on the Majoranas as:
\begin{equation}
	\begin{gathered}
	\psi_j\rightarrow \psi_{j+1}, 1\leq j<L,\\
	\psi_{L}\rightarrow -\psi_1.
\end{gathered}
	\label{}
\end{equation}
This transformation can be implemented by the following unitary operator:
\begin{equation}
	T_y=B_{1,2}\cdots B_{L-1,L},
	\label{}
\end{equation}
where $B_{ij}$ is the exchange operator~\cite{IvanovPRL2001}:
\begin{equation}
	B_{ij}=\frac{1-\psi_i\psi_j}{\sqrt{2}}=e^{-\frac{\pi}{4}\psi_i\psi_j},
	\label{}
\end{equation}
which acts as $B_{ij}\psi_iB_{ij}^\dag = \psi_j, B_{ij}\psi_jB_{ij}^\dag = -\psi_i$. Notice that the overall phase of $T_y$ is ambiguous.

We now evaluate the trace of $T_y$ in the maximally mixed state. Expanding the product of $B$'s, only two terms contribute: the identity, and the total fermion parity, which is from $B_{23}, B_{45}, \cdots, B_{L1}$:
\begin{equation}
	(-1)^{L/2-1}\psi_2\psi_3\cdots \psi_L\psi_1=(-1)^{L/2}\psi_1\psi_2\cdots\psi_L=i^{L/2}.
	\label{}
\end{equation}
Therefore
\begin{equation}
	\begin{split}
	|\langle T_y\rangle|&=\frac{1}{D}|\Tr T_y|\\
	&=\frac{1}{D}\frac{1}{\sqrt{2}^{L-1}}|\Tr (1+i^{L/2})|\\
	&=\frac{1}{\sqrt{2}^{L-1}}|1+i^{L/2}|\\
	&=\frac{1}{\sqrt{2}^{L-1}}2\Big|\cos\frac{\pi L}{8}\Big|\\
	&=\frac{1}{\sqrt{2}^{L-1}}
	\begin{cases}
		0 & L=4(2k+1)\\
		2 & L=8k\\
		\sqrt{2} & L=2(2k+1)
	\end{cases}.
	\end{split}
	\label{eq:eq77}
\end{equation}

Next we consider adding nontrivial dynamics to the boundary theory. The simplest choice is just turning on $-\sum_jS_j$ on the boundary, which after fermionization becomes the free Majorana chain:
\begin{equation}
	H=-\sum_{j=1}^{L-1} i\psi_j\psi_{j+1} + i\psi_L\psi_1.
	\label{}
\end{equation}
Note that the translation action automatically puts the Majorana chain in the sector with an anti-periodic boundary condition. Equivalently, the boundary is described by an Ising CFT projected to the $\Z_2$ symmetric sector. We can directly evaluate the disorder parameter of the translation operator in the continuum limit (see Appendix \ref{app:CFTpartial} for details), which remarkably gives the same $L$ dependence $|2\cos\frac{\pi L}{8}|$ in Eq. \eqref{eq:eq77} from the maximally mixed state. However, we notice that the continuum calculation does not capture the additional $\sqrt{2}$ (from $\sqrt{2}^{L-1}$ in the denominator in Eq. \eqref{eq:eq77}).

Notice that if the region $M$ is a half of a cylinder, we also need to take into acount the physical edge of the cylinder. To lift degeneracy we can turn on a Hamiltonian $-\sum_j S_j$ on the edge.   Assuming $T_y$ is not spontaneously broken, it does not have any nontrivial contribution to the disorder parameter.

\subsubsection{$\Z_N$ plaquette model}
Let us now generalize the result from $\Z_2$ plaquette to $\Z_N$ plaquette model. We will focus on the case with $N$ an odd integer. As already shown earlier, the boundary Hilbert space is a $\Z_N$ spin chain of length $L/2$, projected to the $\Z_N$-invariant subspace. We will denote the (effective) $\Z_N$ spin operators on the boundary by $X_j$ and $Z_j$, and the $\Z_N$ symmetry is generated by $P=\prod_{j=1}^L X_j$. The translation symmetry becomes the Kramers-Wannier duality of the $\Z_N$ spin chain.

Generalizing the derivation in the $N=2$ case, it will be convenient to represent the $\Z_N$ chain in terms of parafermion operators~\cite{Fendley2012}:
\begin{equation}
	\alpha_{2j-1}=Z_j\prod_{k=1}^{j-1} X_k,
	\alpha_{2j}= Z_j\prod_{k=1}^j X_k.
	\label{}
\end{equation}
It is easy to show that $\alpha_j^N=1$. More importantly, they obey non-local commutation relations:
\begin{equation}
	\alpha_i\alpha_j=\omega\alpha_j\alpha_i, 1\leq i<j\leq L.
	\label{}
\end{equation}
We also have
\begin{equation}
	\alpha_{2j-1}^\dag \alpha_{2j}=X_j.
	\label{}
\end{equation}
Thus the total $\Z_N$ charge is given by
\begin{equation}
	P= \prod_{j=1}^{L/2}\alpha_{2j-1}^\dag \alpha_{2j}.
	\label{}
\end{equation}
The Kramers-Wannier duality of the spin chain becomes the translation symmetry of the parafermions:
\begin{equation}
	T_y: \alpha_j\rightarrow \alpha_{j+1}.
	\label{}
\end{equation}

To write down an explicit form for $T$, we define the exchange operator for a pair of parafermion operators. For $i<j$, define a unitary $B_{ij}$ such that~\cite{Clarke2013, LindnerPRX2012}
\begin{equation}
	B_{ij}:\alpha_i\rightarrow\alpha_j, \alpha_j\rightarrow \omega\alpha_i^\dag\alpha_j^2.
	\label{}
\end{equation}
An explicit expression for $B_{ij}$ is given by
\begin{equation}
	B_{ij}=\frac{1}{\sqrt{N}}\sum_{n=0}^{N-1}\omega^{-m(n^2-n)}P_{ij}^n.
	\label{}
\end{equation}
Here $m=\frac{N+1}{2}$.  One can check that $B_{ij}$ preserves the total $\Z_N$ charge $P$. We provide a derivation of $B_{ij}$ in Appendix~\ref{app:exchange}.

With the exchange operator, we can represent the translation $T_y$ as follows:
\begin{equation}
	T_y=B_{12}\cdots B_{L-2,L-1}B_{L-1,L},
	\label{eqn:B2}
\end{equation}
under which
\begin{equation}
	\begin{gathered}
	\alpha_j\rightarrow \alpha_{j+1}, 1\leq j<L,\\
	\alpha_L\rightarrow \omega^2\alpha_1 P^2.
	\end{gathered}
	\label{}
\end{equation}
So with a fixed $P$, the unitary Eq. \eqref{eqn:B2} indeed acts as translation.

Now we are ready to calculate the disorder operator:
\begin{equation}
	\begin{split}
		|\langle T_y\rangle|&=\frac{1}{\sqrt{N}^{L-1}}\frac{1}{D}\big|\Tr \sum_{n=0}^{N-1}P^n \omega^{-m(n^2-n)L/2}\big|\\
		&=\frac{1}{\sqrt{N}^{L-1}} \big|\sum_{n=0}^{N-1} \omega^{-m(n^2-n)L/2}\big|\\
		&=\frac{1}{\sqrt{N}^{L-1}} \big|\sum_{n=0}^{N-1} \omega^{-mn^2L/2}\big|\\
		&=\frac{1}{\sqrt{N}^{L-1}}\sqrt{N\cdot \mathrm{gcd}\left( \frac{mL}{2},N \right)}.
	\end{split}
	\label{}
\end{equation}

For $N$ an odd prime, we find
\begin{equation}
	|\langle T_y\rangle|=\frac{1}{\sqrt{N}^{L-1}}
	\begin{cases}
		N & L\equiv 0\:(\text{mod }N)\\
		\sqrt{N} & \text{otherwise}
	\end{cases}.
	\label{}
\end{equation}

\subsubsection{DMRG results}
\label{sec:IVC2}

\begin{figure}[!t]
\includegraphics[angle=0,width=\linewidth]{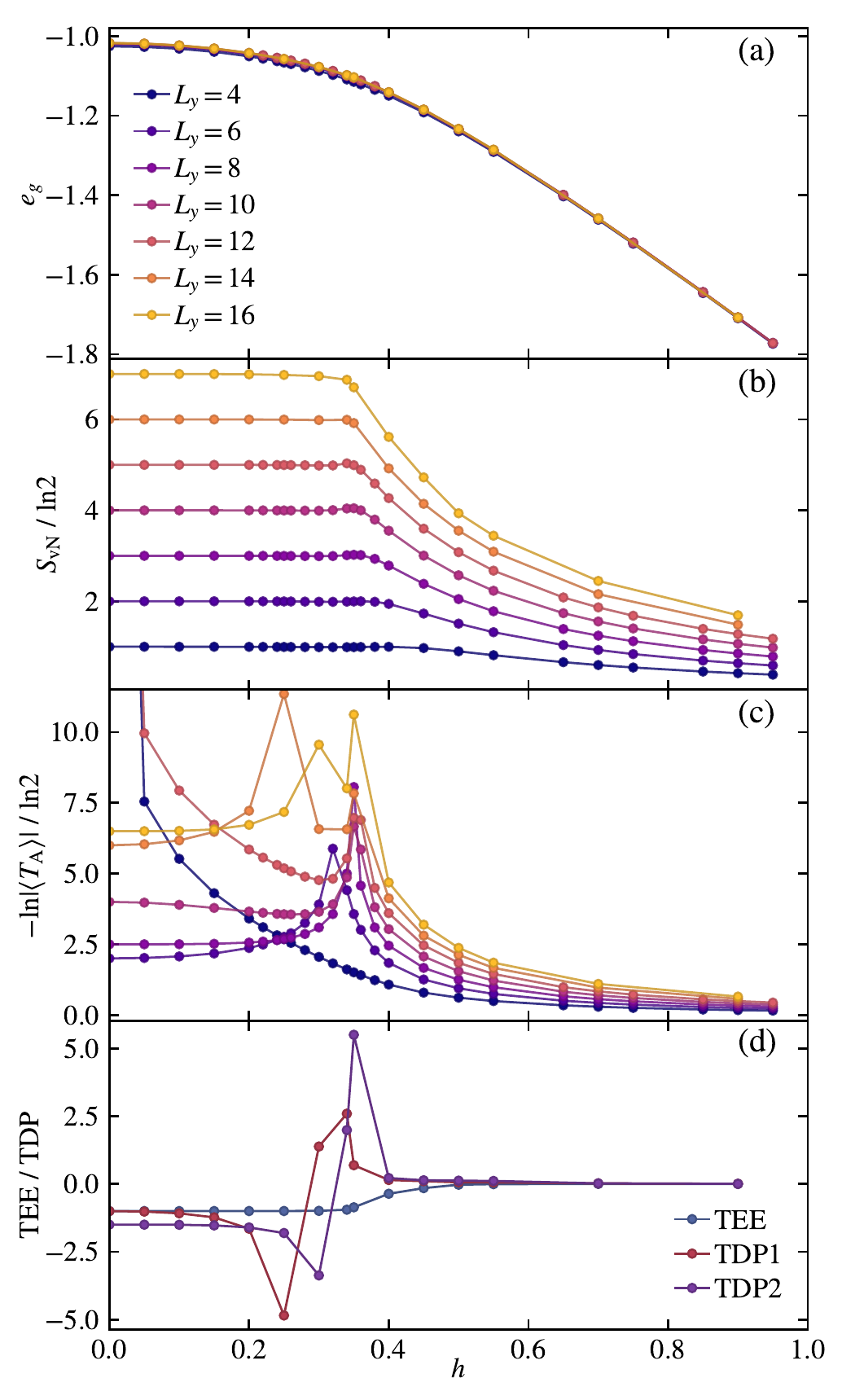}
\caption{For Wen's plaquette model on cylinder of various circumference $L_y$, (a) ground-state energy $e_g$, (b) von Neumann entanglement entropy (EE) $S_{\mathrm{vN}}$ and (c) negative logarithmic partial translation expectation value $-\ln{|\langle T_{\rm A}\rangle|}$, are shown versus the magnetic field $h$. Panel (d) shows the topological entanglement entropy (TEE), and the topological disorder parameter extrapolated from $-\ln{|\langle T_{\rm A}\rangle|}$ for $L_y=6,10,14$ (denoted as TDP1) and the one for $L_y=8,16$ (denoted as TDP2), in units of $\ln{2}$. }
\label{Fig:WenPlaqPhase}
\end{figure}

\begin{figure}[!t]
\includegraphics[angle=0,width=\linewidth]{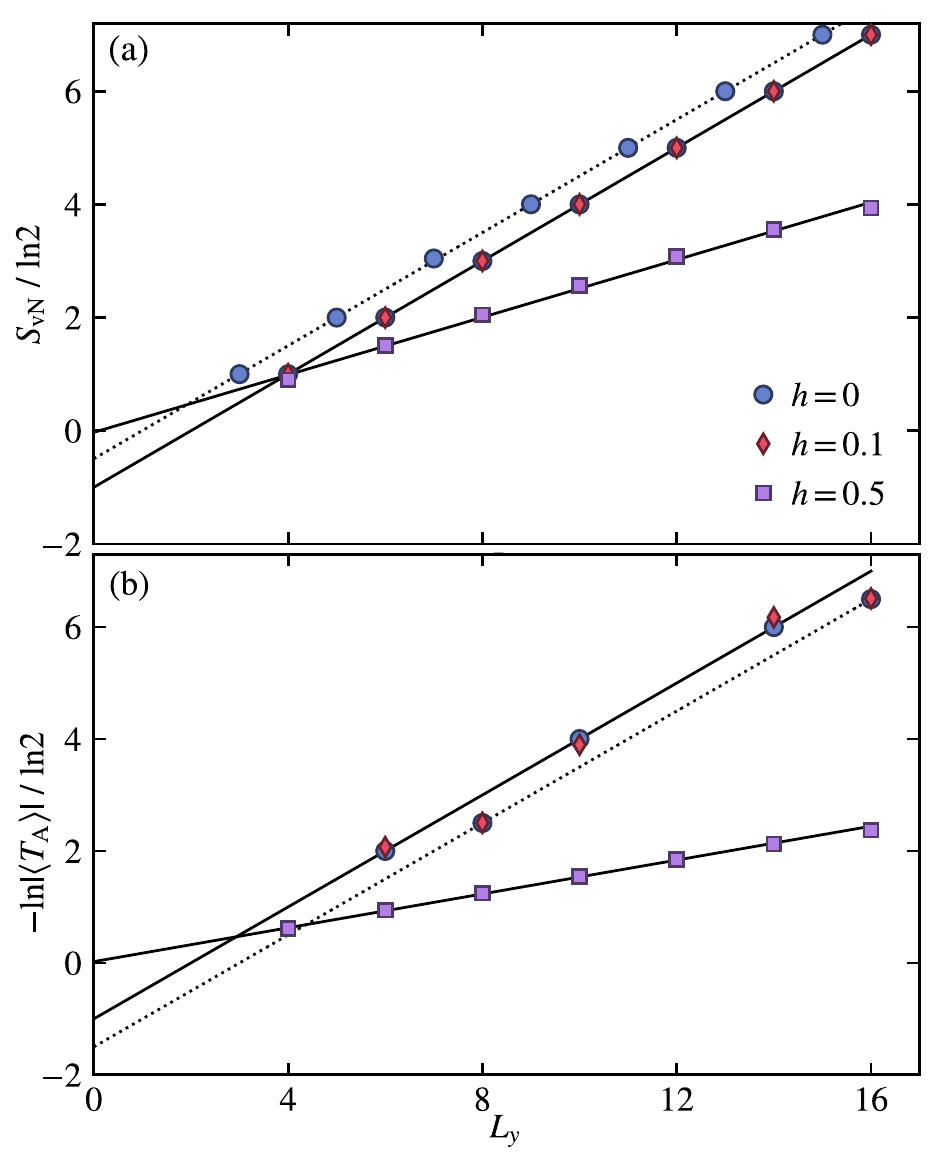}
\caption{For Wen's plaquette model on cylinder with three different fields $h=0, 0.1, 0.5$,
(a) entanglement entropy (EE) is shown versus circumferences $L_y$.
For the unperturbed case $h=0$, EE's are well extrapolated to $-\ln{2}$ ($-\ln\sqrt{2}$)
for even (odd) circumferences, as $L_y\rightarrow0$.
For $h=0.1$ and $h=0.5$, only even $L_y$ is considered and the corresponding EE's are extrapolated to $-\ln{2}$ and $0$.
(b) The negative logarithmic value of DP,
$ -\ln|{\langle T_\mathrm{A} \rangle}|$,
is shown versus circumferences $L_y$.
For $h=0$, it are extrapolated to $-\ln{2}$ for $L_y=6,10,14$, and to $-\frac{3}{2}\ln{2}$ for $L_y=8,16$.
For $h=0.5$, DP extrapolates to $0$.}
\label{Fig:WenPlaqTDP}
\end{figure}

Similar as in Sec.~\ref{sec:IVA2}, for the system away from exactly solvable limit and perturbative regime,
we perform DMRG simulations to compute the TDP of Wen's plaquette model
with both transverse and longitudinal fields in the Hamiltonian in Eq.~\eqref{eq:eq61}.
In the DMRG calculations, we fix the length of the cylinder to $L_x=16$, and vary the circumference from $L_y=4$ to $L_y=16$,
with maximal $D=2048$ bond states kept which ensures sufficiently small truncation errors $\epsilon\sim10^{-5}$.

The numerical results are shown in \Fig{Fig:WenPlaqPhase} and \Fig{Fig:WenPlaqTDP}. The ground-state energy $e_g$, entanglement entropy (EE) $S_{\mathrm{vN}}$, the disorder operator $\langle T_A \rangle$ for partial translation, and their extrapolated values are shown
in Fig.~\ref{Fig:WenPlaqPhase} (a-d), respectively.
To examine the topological order in the small-$h$ cases, we consider the finite size scaling of entanglement entropy (EE) data in Fig.~\ref{Fig:WenPlaqPhase} (b) and Fig.~\ref{Fig:WenPlaqTDP}.
In \Fig{Fig:WenPlaqTDP}(a), EE's  versus $L_y$ are shown for $h=0$ and $h=0.1$, and the data are well extrapolated to $-\ln{2}$ ($-\ln\sqrt{2}$)  for even (odd) circumference as $L_y\rightarrow0$. That is, the system possesses a finite topological entanglement entropy (TEE) for those $h$'s, confirming their topological ordered nature.
On the other hand, the EE data for $h=0.5$ are extrapolated to $0$, as expected for the topologically trivial phase.
In \Fig{Fig:WenPlaqPhase}(d), we have performed such extrapolation for all the $h$'s concerned and for the even circumference cases.
It shows that at around $h\simeq h_c$ TEE undergoes a change from $-\ln{2}$ to $0$.
We also note that, when performing the EE calculations on odd-circumference cylinders for small $h$,
the EE data are extrapolated to $-\ln\sqrt{2}$ instead of $-\ln{2}$. Physically, this is because an odd-circumference cylinder is in the EMD defect sector, so the TEE increases by $\ln \sqrt{2}$.

The results of DP and  TDP for partial translation are more intricate.
For TDP, we find that it clearly vanishes in the trivial phase
[c.f. \Fig{Fig:WenPlaqTDP}(b)].
However, inside the topological phase, one needs to classify \{$L_y$\} into 3 classes, $L_y=4(2k+1),~ 8k,~ 2(2k+1)$ with $k\in\Z$.
For the first class of $L_y=4(2k+1)$,
i.e. circumference being odd multiples of 4, we always get zero values for TDP in the vicinity of $h=0$. For the second (third) cases, TDPs have finite values, which are extrapolated to $\frac{3}{2}\ln2$ ($\ln{2}$) as $L_y\rightarrow0$
as shown in \Fig{Fig:WenPlaqTDP}(b). These results are fully consistent with  the discussion in Sec.~\ref{sec:IVC1}, in particular, Eq.~\eqref{eq:eq77}, as well as that in Appendix.~\ref{app:CFTpartial}.
In \Fig{Fig:WenPlaqPhase}(d),
we show the extrapolated values of TDP (TDP1 for $L_y=2(2k+1)$ and TDP2 for $L_y=8k$) as a function of $h$. 
It is clearly seen that, $\gamma\simeq\ln{2}$ for the TDP1 cases and
$\gamma\simeq\frac{3}{2}\ln{2}$ for the TDP2 cases in the vicinity of $h=0$. In Fig.~\ref{Fig:WenPlaqPhase} (c), we also note that as a function of $h$ the (logarithmic) disorder parameters always show peaks around the transition point $h_c$.

\section{Conclusion and Discussion}
\label{Sec:V}
In this work we introduce a new topological invariant for (2+1)d gapped phases with global symmetry. We show that the ground state expectation value of the disorder operator for a connected region $M$ exhibits the following scaling form:
\begin{equation}
	|\langle U_M(g) \rangle|\approx d_g e^{-\alpha |\partial M|},
	\label{eqn:Ug_one_more_time}
\end{equation}
where $d_g$ is a quantized invariant determined by both the quantum dimensions of $U(g)$ symmetry defects as well as their local degeneracy. When the entanglement Hamiltonian can be approximated by a (1+1)d CFT, we derive a precise formula for $d_g$. We also study a wide range of examples, in particular in lattice models such as $\Z_N$ toric code and Wen's plaquette models, with both CFT and non-CFT entanglement Hamiltonian, to demonstrate the validity of Eq. \eqref{eqn:Ug_one_more_time} and the relation between $d_g$ with quantum dimensions of $U_g$ defects.

In all our calculations we have reduced the disorder parameter to the thermal expectation value of a global symmetry transformation in a (1+1)d system in the high temperature limit. Therefore, our result can also be interpreted as a universal invariant of the (1+1)d system with global symmetry. It is an interesting question to establish the result directly in a (1+1)d theory, especially beyond CFT. 

In this work we focus on (2+1)d gapped phases as the symmetry defects are well-understood. It will be interesting to understand what happens in higher dimensions. For instance, the quantum double model studied in can be easily generalized to arbitrary dimensions and in fact the result does not really depend on spatial dimension. There are also generalizations of electromagnetic duality symmetry in higher dimensions, such as the duality group of U(1) gauge theory in (3+1)d or $\Z_N$ 2-form gauge theory in (4+1)d~\cite{Chen21}. Disorder parameters in these examples should also reveal topological corrections related to quantum dimensions of defects. Like the (2+1)d case, the disorder parameter can be expressed as a thermal average of the global symmetry transformation in the entanglement Hamiltonian. Assuming that the entanglement Hamiltonian is qualitatively similar to the boundary Hamiltonian, similar questions can be raised for ``duality'' symmetry in the boundary theory, when there is self duality.

\section*{Acknowledgement}
M.C. would like to thank Wenjie Ji for helpful discussions about Ref. [\onlinecite{Ji2021}].
B.B.C. and Z.Y.M. would like to thank Zheng Yan and Jiarui Zhao for inspiring discussions and collaborations on related topics and they acknowledge support from the RGC of Hong Kong SAR of China (Grant Nos.  17303019, 17301420, 17301721 and AoE/P-701/20), the K. C. Wong
Education Foundation (Grant No. GJTD-2020-01) and
the Seed Funding ``Quantum-Inspired explainable-AI" at
the HKU-TCL Joint Research Centre for Artificial Intelligence. H.H.T. acknowledges support from the Deutsche Forschungsgemeinschaft (DFG) through project A06 of SFB 1143 (project-id 247310070). M.C. acknowledges support from NSF under award number DMR-1846109. We thank the Computational Initiative at the Faculty of Science and the Information Technology Services at the University of Hong Kong and the Tianhe platforms at the National Supercomputer Center in Guangzhou for their technical support and generous allocation of CPU time.

\appendix

\section{TDP on a cylinder}
\label{sec:TDP_on_cylinder}
In this Appendix we calculate TDP when the region is a half of the cylinder, with boundary along the periodic direction. We will also assume that there is an anyon flux $a$ through the cylinder. Apparently $a$ has to be invariant under the symmetry, otherwise the disorder parameter vanishes. The reduced density operator is given by
\begin{equation}
	\rho=\frac{1}{\cal{Z}_a}e^{-\frac{\xi_l}{l} H_l}e^{-\frac{\xi_r}{l}H_r}.
\end{equation}
Here $\xi_l$ and $\xi_r$ are the effective correlation length on the left and right edges.
Importantly, the left (physical) edge has $\xi_l=\infty$, while the right edge, which is the entanglement cut, is at a finite but high temperature $\xi_r$~\cite{TuPRB2013}. We note a similar geometry has been used in the computation of entanglement spectrum in (2+1)d quantum many-body systems~\cite{ZhengYan2022ES}. The disorder parameter is then given by
\begin{equation}
	\langle U_g(M)\rangle = \frac{\Tr_{\cal{H}_{\bar{a}}} U_g e^{-\frac{\xi_l}{l} H_l} \Tr_{\cal{H}_a} U_g e^{-\frac{\xi_r}{l} H_r}}{\Tr_{\cal{H}_{\bar{a}}}  e^{-\frac{\xi_l}{l} H_l} \Tr_{\cal{H}_a}  e^{-\frac{\xi_r}{l} H_r}}.
\end{equation}
The left edge, being at the zero temperature, is dominated by the ground state contribution in the sector $\bar{a}$. Since the sector $\mathcal{H}_{\bar{a}}$ is invariant under the transformation, we may assume that the symmetry acts on the highest weight states as a unitary matrix, whose trace is $\chi_{\bar{a}}(g)$:
\begin{equation}
	\frac{\Tr_{\cal{H}_{\bar{a}}} U_g e^{-\frac{\beta_l}{l}H_l} }{\Tr_{\cal{H}_{\bar{a}}}  e^{-\frac{\xi_l}{l}H_l} } \approx \frac{\chi_{\bar{a}}(g)}{p_{\bar{a}}(0)},
\end{equation}
Here $p_a(0)$ is the degeneracy of the highest weight space.

For the right entanglement edge, we again use modular transformation to evaluate the partition function:
\begin{equation}
	\begin{split}
		\mathcal{Z}_{a}^{(\id,g)}\left( \frac{i\xi_r}{l} \right)&=\Tr_{\mathcal{H}_a} U_g e^{-\frac{\xi_r}{l} H_r}\\
		&=\sum_{b_g} \mathcal{S}_{a,b_g}^{(\id,g)}\mathcal{Z}_{b_g}^{(g,\id)}\Big(\frac{il}{\xi_r}\Big)\\
		&\approx \left(\sum_{b_g\in \Lambda_g}\mathcal{S}_{a,b_g}^{(\id,g)} p_{b_g}\right) e^{-\frac{2\pi h_g}{\xi_r}l},
	\end{split}
	\label{}
\end{equation}
and the denominator
\begin{equation}
	\begin{split}
		\Tr_{\mathcal{H}_a} e^{-\frac{\xi_r}{l}H_r} &=\sum_{b} S_{ab}\mathcal{Z}_{b}\Big(\frac{il}{\beta_r}\Big)\approx S_{a0}\\
	\end{split}
	\label{}
\end{equation}
We thus find
\begin{equation}
	\langle U_g(M)\rangle=\frac{\chi_{\bar{a}}(g)}{p_{\bar{a}}(0)}
	\left(\sum_{b_g\in \Lambda_g}\frac{\mathcal{S}_{a,b_g}^{(\id,g)}}{S_{a0}} p_{b_g}\right) e^{-\frac{2\pi h_g}{\xi_r}l}.
\end{equation}
\section{Quantum dimension of genons}
\label{app:genons}
We compute the dimension of $0_R$ directly from the fusion rule using Verlinde formula:
\begin{equation}
\begin{split}
    d_{0_R}^2&=\sum_{a_1,\cdots,a_n}N_{a_1a_2\cdots a_n}^0 d_{a_1}d_{a_2}\cdots d_{a_n}\\
    &=\sum_{a_1,\cdots,a_n} \sum_{x}\frac{S_{a_1x}S_{a_2x}\cdots S_{a_nx}}{S_{0x}^{n-2}} d_{a_1}d_{a_2}\cdots d_{a_n}\\
    &= \sum_xS_{0x}^{2-n} \left( \sum_a d_a S_{ax}\right)^n\\
    &=\sum_xS_{0x}^{2-n}\cal{D}^n \left( \sum_a S_{0a}S_{ax}\right)^n\\
    &=S_{00}^{2-n}\cal{D}^n\\
	&=\cal{D}^{2n-2}.
\end{split}
\end{equation}

\section{Exchange operator for parafermions}
\label{app:exchange}
For two $\Z_N$ parafermions $\alpha_1$ and $\alpha_2$, suppose they satisfy $\alpha_1\alpha_2=\omega\alpha_2\alpha_1$. Assume $N$ is an odd integer. Define the $\Z_N$ charge $P_{12}=\alpha_1^\dag\alpha_2$, and consider unitary operators of the form~\cite{LindnerPRX2012}
\begin{equation}
	B_{12}=\frac{1}{\sqrt{N}}\sum_{n=0}^{N-1}\omega^{-mn^2+qn}P_{12}^n.
	\label{}
\end{equation}
First we prove $B_{12}$ is a unitary. We can work in the eigenbasis of $P_{12}$, setting $P_{12}=\omega^k$,
\begin{equation}
	\begin{split}
		B_{12}&=\frac{1}{\sqrt{N}}\sum_{n=0}^{N-1}\omega^{-mn^2+(k+q)n}\\
		&=\frac{1}{\sqrt{N}}\sum_{n=0}^{N-1}\omega^{-m[n^2-m^{-1}(k+q)n]}\\
		&=\frac{1}{\sqrt{N}}\sum_{n=0}^{N-1}\omega^{-m[n-2^{-1}m^{-1}(k+q)]^2}\\
		&=\omega^{a(k+q)^2}\frac{1}{\sqrt{N}}\sum_{n=0}^{N-1}\omega^{-mn^2}
	\end{split}
	\label{}
\end{equation}
Since $a=(4m)^{-1}$, i.e. $4ma\equiv 1\,(\mathrm{mod }N)$. The remaining Gauss sum can be evaluated in closed form:
\begin{equation}
	g_m\equiv\frac{1}{\sqrt{N}}\sum_{n=0}^{N-1}\omega^{-mn^2}=\varepsilon_N \left( \frac{m}{N} \right).
	\label{}
\end{equation}
where
\begin{equation}
	\varepsilon_N=
	\begin{cases}
		1 & N\equiv 1 \text{ mod }4\\
		i & N\equiv 3 \text{ mod }4
	\end{cases}.
	\label{}
\end{equation}
It is sufficient for our purpose to know that the Gauss sum evaluates to a phase factor. Thus we have shown that $B_{12}$ is a unitary.

Now we compute $B_{12}\alpha_1 B_{12}^\dag$. First we notice
\begin{equation}
	P_{12}\alpha_1=\alpha_1^\dag\alpha_2\alpha_1=\omega^{-1}\alpha_1P_{12}.
	\label{}
\end{equation}
Therefore we have
\begin{equation}
	\begin{split}
		B_{12}\alpha_1B_{12}^\dag &= \frac{1}{\sqrt{N}}\sum_{n=0}^{N-1}\omega^{-mn^2+qn}P_{12}^n \alpha_1 B_{12}^\dag\\
		&=\alpha_1\frac{1}{\sqrt{N}}\sum_{n=0}^{N-1}\omega^{-mn^2+qn}\omega^{-n}P_{12}^n  B_{12}^\dag\\
		&=\alpha_1\frac{1}{\sqrt{N}}\sum_{n=0}^{N-1}\omega^{-mn^2+qn}\omega^{(k-1)n}  B_{12}^\dag\\
		&=\alpha_1 |g_m|^2 \omega^{a(k+q-1)^2} \omega^{-a(k+q)^2}\\
		&=\alpha_1 \omega^{a[2(k+q)-1]}\\
		&=\omega^{a(2q-1)}\alpha_1P_{12}^{2a}
	\end{split}
	\label{}
\end{equation}
Set $q=m=2^{-1}=\frac{N+1}{2}$, we obtain
\begin{equation}
	B_{12}\alpha_1B_{12}^\dag=\alpha_1P_{12}=\alpha_2.
	\label{}
\end{equation}

\onecolumngrid
\section{CFT analysis}
\label{app:CFTpartial}

\subsection{Ising and Spin$(\nu)_1$ CFTs}
In the following $\tau$ denotes the complex parameter of a 2D torus, and $q=e^{2\pi i\tau}$. First define the partition functions for a free Majorana fermion
\begin{equation}
	\begin{split}
		Z_\mathrm{AA}(\tau)&=\Tr_{\mathrm{NS}}q^{L_0-\frac{1}{48}}=q^{-1/48}\prod_{n=0}^\infty (1+q^{n+1/2}),\\
		Z_\mathrm{AP}(\tau)&=\Tr_{\mathrm{NS}}(-1)^{N_f}q^{L_0-\frac{1}{48}}=q^{-1/48}\prod_{n=0}^\infty (1-q^{n+1/2}),\\
		Z_\mathrm{PA}(\tau)&=\frac{1}{\sqrt{2}}\Tr_{\mathrm{R}}q^{L_0-\frac{1}{48}}=\frac{q^{1/24}}{\sqrt{2}}\prod_{n=0}^\infty (1+q^n).
	\end{split}
	\label{}
\end{equation}
Here P/A means periodic/anti-periodic boundary condition in the spatial or temporal direction.

The chiral Majorana theory is closely related to the Ising CFT. In fact, the latter can be obtained from coupling the Majorana fermion to a $\Z_2$ gauge field~\footnote{ Technically, applying the Gliozzi-Scherk-Olive (GSO) projection.}. More generally, by coupling $\nu$ copies of chiral Majorana fermions one obtains the Spin$(\nu)_1$ CFT.

When $\nu$ is odd, Spin$(\nu)_1$ has three primaries, which will be labeled as $1,\psi$ and $\sigma$. The corresponding characters are
\begin{equation}
	\begin{split}
		\chi_1^{(\nu)}=\frac12(Z_{\text{AA}}^\nu+Z_{\text{AP}}^\nu),
		\chi_{\psi}^{(\nu)}=\frac12(Z_{\text{AA}}^\nu-Z_{\text{AP}}^\nu),
	\chi_{\sigma}^{(\nu)}=\frac{1}{\sqrt{2}}Z_\mathrm{PA}^\nu.
	\end{split}
	\label{}
\end{equation}
Note that the $\nu=1$ case is actually the Ising CFT. We will suppress the superscript in this case, i.e $\chi_a^{(1)}\equiv \chi_a$ for $a=1,\psi,\sigma$.

The corresponding conformal dimensions are $h_1=0,h_\psi=\frac12, h_\sigma=\frac{\nu}{16}$. The modular S and T matrices of the theory read:
\begin{equation}
	S=\frac12
	\begin{pmatrix}
		1 & 1 & \sqrt{2}\\
		1 & 1 & -\sqrt{2}\\
		\sqrt{2} & -\sqrt{2} & 0
	\end{pmatrix},
	T_{ab}=\delta_{ab}e^{2\pi i (h_a-\frac{c}{24})}.
	\label{}
\end{equation}

For $\nu$ even, Spin$(\nu)_1$ has four primaries, labeled as $1,\psi,v_+,v_-$, with conformal dimensions $h_1=0,h_\psi=\frac12, h_{v_\pm}=\frac{\nu}{8}$. The characters are
\begin{equation}
	\chi_1^{(\nu)}=\frac12(Z_\text{AA}^\nu+Z_\text{AP}^\nu),
	\chi_{\psi}^{(\nu)}=\frac12(Z_\text{AA}^\nu-Z_\text{AP}^\nu),
	\chi_{v_\pm}^{(\nu)}=\frac{1}{{2}}Z_\mathrm{PA}^\nu.
	\label{}
\end{equation}

\subsection{EDM disorder parameter}
At the level of bulk topological order, Spin($\nu)_1$ can be viewed as a $\Z_2$ gauge theory coupled to fermionic matter. For even $\nu$, the two fermion parity vortices $v_\pm$ are completely symmetric and there is a $\Z_2$ symmetry that swaps the two. In the chiral CFT, such a symmetry can be realized as the fermion parity of one of the chiral Majorana fermion. Without loss of generality, let us choose it to be $(-1)^{N_1}$. We now calculate the disorder parameter for this symmetry.

\begin{equation}
		\begin{split}
			\langle (-1)^{N_1}\rangle&=\frac{1}{Z}\Tr_\text{NS}\frac{1+(-1)^{N_f}}{2}(-1)^{N_1} q^{L_0-\frac{c}{24}}\\
			&= \frac{\frac12[Z_\text{AA}(\tau)Z_\text{AP}(\tau)^{\nu-1}+Z_\text{AP}(\tau)Z_\text{AA}(\tau)^{\nu-1}]}{\chi_1(\tau)}\\
			&=\frac{\chi_1(\tau) \chi_1^{(\nu-1)}(\tau) - \chi_\psi(\tau) \chi_\psi^{(\nu-1)}(\tau)}{\chi_1^{(\nu)}(\tau)}
	\end{split}
	\label{eqn:chiralPf1}
\end{equation}

Now for a purely imaginary $\tau=\frac{i\beta}{l}$, we use modular transformations to find the asymptotic forms of the characters for small $\beta\ll l$:
\begin{equation}
	\chi_a\left( \frac{i\beta}{l} \right)=\sum_bS_{ab}\chi_b\left(\frac{il}{\beta}\right).
	\label{}
\end{equation}
Then we can expand the character
\begin{equation}
	\chi_b\left( \frac{il}{\beta} \right)=\sum_{m=0}^\infty p_b(m)e^{-\frac{2\pi l}{\beta}(h_b+m-\frac{c}{24})}\approx p_b(0)e^{-\frac{2\pi l}{\beta}(h_b-\frac{c}{24})}.
	\label{eqn:chi-expansion}
\end{equation}

Applying the approximation to Eq. \eqref{eqn:chiralPf1} and keeping only the most relevant terms, we find
\begin{equation}
	\begin{split}
	\langle (-1)^{N_1}\rangle
	&\approx{\sqrt{2}} \frac{\chi_\sigma(\frac{il}{\beta})\chi_1^{(\nu-1)}(\frac{il}{\beta})+\chi_\sigma^{(\nu-1)}(\frac{il}{\beta})\chi_1(\frac{il}{\beta})}{\chi_1^{(\nu)}(\frac{il}{\beta})}\\
&\approx
\begin{cases}
	2\sqrt{2}e^{-\frac{\pi l}{8\beta}} & \nu=2\\
	\sqrt{2}e^{-\frac{\pi l}{8\beta}} & \nu>2
\end{cases}.
	\end{split}
	\label{}
\end{equation}

We now turn to a different but closely related example, that is an internal EDM symmetry in a $\Z_2$ toric code.
When the symmetry is present, the boundary of the $\Z_2$ toric code must be gapless. One family of possible boundary theories is the Spin$(2n+1)_1$ CFTs, and the $n=0$ case is the Ising CFT. These CFTs can all be represented as (non-chiral) Majorana fermions coupled to a $\Z_2$ gauge field. The EDM is realized as ``chiral'' fermion parity, say $(-1)^{N_R}$.

First let us write down the vacuum character for the CFT, from GSO projection of the Majorana fermions:
\begin{equation}
	Z_1=\Tr_\text{NS} \frac{1+(-1)^{N_L+N_R}}{2} q^{L_0-\frac{1}{48}}\ol{q}^{\ol{L}_0-\frac{1}{48}}=\frac12(|\chi_1^{(\nu)}(\tau)|^2+|\chi_\psi^{(\nu)}(\tau)|^2).
	\label{}
\end{equation}
To calculate the disorder parameter for $(-1)^{N_R}$, we need
\begin{equation}
	\Tr_\text{NS} \frac{1+(-1)^{N_L+N_R}}{2} (-1)^{N_R}q^{L_0-\frac{1}{48}}\ol{q}^{\ol{L}_0-\frac{1}{48}}=\frac12(|\chi_1^{(\nu)}(\tau)|^2-|\chi_\psi^{(\nu)}(\tau)|^2).
	\label{}
\end{equation}

Putting together we have
\begin{equation}
	\begin{split}
		\langle (-1)^{N_R}\rangle &= \frac{|\chi_1^{(\nu)}(\frac{i\beta}{l})|^2 - |\chi_\psi^{(\nu)}(\frac{i\beta}{l})|^2}{|\chi_1^{(\nu)}(\frac{i\beta}{l})|^2+|\chi_\psi^{(\nu)}(\frac{i\beta}{l})|^2}\\
		&\approx\sqrt{2}\,\frac{\chi_1^{(\nu)}(\frac{il}{\beta})\ol{\chi_\sigma^{(\nu)}(\frac{il}{\beta})}+\text{c.c.}}{|\chi_1^{(\nu)}(\frac{il}{\beta})|^2}\\
		&\approx 2^{n+1} \sqrt{2}e^{-\frac{\pi l}{8\beta}}.
	\end{split}
	\label{}
\end{equation}
Here we used the fact that $p_\sigma(0)=2^n$ (i.e. the dimension of the spinor representation) for Spin($2n+1)_1$ CFT.

Next we consider what happens if velocity anisotropy between the $2n+1$ Majorana fermions is introduced, so the boundary theory is no longer a CFT. In the following denote $\nu=2n+1$. The entanglement Hamiltonian is now assumed to be
\begin{equation}
	H_E=\sum_{i=1}^\nu\beta_i \left(L_0^{(i)}+\ol{L}_0^{(i)}-\frac{1}{24}\right),
	\label{}
\end{equation}
where $L_0^{(i)}$ is the Hamiltonian for the $i$-th chiral Majorana fermion.
\begin{equation}
		\begin{split}
			\langle (-1)^{N_1}\rangle &= \frac{\prod_{i=1}^\nu Z_\text{AA}(\tau_i)\prod_{i=1}^\nu \ol{Z_\text{AP}(\tau_i)}+\prod_{i=1}^\nu Z_\text{AP}(\tau_i)\prod_{i=1}^\nu \ol{Z_\text{AA}(\tau_i)}}{\prod_{i=1}^\nu |Z_\text{AP}(\tau_i)|^2+\prod_{i=1}^\nu |Z_\text{AA}(\tau_i)|^2}\\
	\end{split}
\end{equation}
Under modular S transformation,
\begin{equation}
	Z_\text{AA}(\tau)=Z_\text{AA}(-1/\tau), Z_\text{AP}(\tau)=Z_\text{PA}(-1/\tau).
	\label{}
\end{equation}
Plug in $\tau_i=\frac{i\beta_i}{l}$, we have
\begin{equation}
		\begin{split}
			\langle (-1)^{N_R}\rangle &= \frac{\prod_{i=1}^\nu Z_\text{AA}(\frac{il}{\beta_i})\prod_{i=1}^\nu \ol{Z_\text{PA}(\frac{il}{\beta_i})}+\prod_{i=1}^\nu Z_\text{PA}(\frac{il}{\beta_i})\prod_{i=1}^\nu \ol{Z_\text{AA}(\frac{il}{\beta_i})}}{\prod_{i=1}^\nu |Z_\text{PA}(\frac{il}{\beta_i})|^2+\prod_{i=1}^\nu |Z_\text{AA}(\frac{il}{\beta_i})|^2}\\
	\end{split}
\end{equation}
Use $Z_\text{AA}=\chi_0+\chi_\psi, Z_\text{PA}=\sqrt{2}\chi_\sigma$ and the expansion Eq. \eqref{eqn:chi-expansion}, we find
\begin{equation}
	\langle (-1)^{N_R}\rangle \approx 2^{n+1}\sqrt{2}e^{-\frac{\pi l}{8}\sum_{i=1}^\nu \frac{1}{\beta_i}}.
	\label{}
\end{equation}
So the TDP is not affected.

\subsection{Partial translation}
We now calculate the disorder parameter for partial translation in $\Z_2$ toric code, assuming that the boundary theory is an Ising CFT. First we need to understand how the lattice translation is represented in the field theory. After fermionization, the Hamiltonian of a critical Majorana chain of length $L$ reads
\begin{equation}
	H=\sum_k \sin k \psi_k^\dag\psi_k-E_0,
	\label{}
\end{equation}
where $k=\frac{2n+1}{L}\pi$ for $n=0,1,\dots, \frac{L}{2}-1$ for NS boundary condition, and $E_0=\frac{1}{2\sin\frac{\pi}{L}}$. We now define $\psi_{Lk}\simeq\psi_k, \psi_{Rk}\simeq\psi_{\pi-k}$ for small $k$, then
at low energy the Majorana fermion theory can be approximated by
\begin{equation}
	H=\sum_{k}k(\psi_{Lk}^\dag \psi_{Lk}+\psi_{Rk}^\dag \psi_{Rk})- E_0,
	\label{}
\end{equation}
where the constant $E_0=\frac{L}{2\pi}+\frac{\pi}{12L}+O(\frac{1}{L^3})$.

The translation operator $T$ acts on the lattice Majorana operators as $T: \psi_j\rightarrow \psi_{j+1}$. In the momentum space, $T$ becomes
\begin{equation}
	\psi_{Lk}\rightarrow e^{ik}\psi_{Lk}, \psi_{Rk}\rightarrow -e^{-ik}\psi_{Rk}.
	\label{}
\end{equation}
which can be more compactly written as
\begin{equation}
	T=(-1)^{N_R}e^{i\frac{P}{L}}.
	\label{}
\end{equation}
Here $P=L_0-\ol{L}_0$ is the CFT momentum.

Thus we have
\begin{equation}
	\begin{split}
		\langle T\rangle=\frac{1}{Z}\Tr_\text{NS} &\frac{1+(-1)^{N_L+N_R}}{2}T e^{2\pi (i\tau_1P-\tau_2 H)} = \frac{|\chi_1\left(\tau+\frac{1}{L}\right)|^2 - |\chi_\psi\left(\tau+\frac{1}{L}\right)|^2}{|\chi_1(\tau)|^2+|\chi_\psi(\tau)|^2}.
	\end{split}
	\label{}
\end{equation}

Now for a purely imaginary $\tau$ with $\text{Im}\, \tau>0$, we use modular transformations to find the asymptotic forms of the characters for small $\beta$~\cite{ShiozakiPRB2017}:
\begin{equation}
	\chi_a\left(\tau+\frac{1}{L}\right)=\sum_b S_{ab}\chi_b\left(-\frac{1}{\tau+\frac{1}{L}}\right)=\sum_b (ST^L)_{ab}\chi_b\left(\frac{\tau L}{\tau+\frac{1}{L}}\right) = \sum_b (ST^LS)_{ab}\chi_b\left(-\frac{1}{L}-\frac{1}{L^2\tau}\right)
	\label{}
\end{equation}
Therefore
\begin{equation}
	\chi_a\left( \frac{i\beta}{L}+\frac{1}{L} \right)=\sum_b(ST^LS)_{ab}\chi_b\left( -\frac{1}{L}+\frac{i}{\beta L} \right),
	\label{}
\end{equation}
and then after expanding the character we have
\begin{equation}
	\chi_b\left( \frac{i}{\beta L}-\frac{1}{L} \right)\simeq e^{\frac{2\pi i}{L}(h_b-\frac{c}{24})}e^{-\frac{2\pi}{\beta L}(h_b-\frac{c}{24})}.
	\label{}
\end{equation}
Note that the expansion is only valid for $\beta L\ll 1$.

We have a similar estimate for the denominator:
\begin{equation}
	\chi_a\left( \frac{i\beta}{L} \right)=\sum_bS_{ab}\chi_b\left(\frac{iL}{\beta}\right)\approx \sum_b S_{ab}e^{-\frac{2\pi L}{\beta}(h_b-\frac{c}{24})},
	\label{}
\end{equation}
for $L/\beta \gg 1$, which is obviously satisfied if $\beta L\ll 1$.

In each case, the leading term is $b=1$ with $h_1=0$, so we finally obtain
\begin{equation}
	\langle T\rangle=\frac{|(S T^L S)_{11}|^2-|(ST^LS)_{\psi 1}|^2}{|S_{11}|^2+|S_{\psi 1}|^2}e^{-\frac{\pi}{12\beta}(L-\frac{1}{L})}.
	\label{}
\end{equation}
The modular transformation $ST^n S$ takes the following form:
\begin{equation}
	ST^n S=e^{-\frac{\pi i n}{24}}\frac{1}{4}\begin{pmatrix}
		1+(-1)^n+2e^{\frac{\pi i n}{8}} & 1+(-1)^n-2e^{\frac{\pi i n}{8}} & \sqrt{2}[1-(-1)^n]\\
		1+(-1)^n-2e^{\frac{\pi i n}{8}} & 1+(-1)^n+2e^{\frac{\pi i n}{8}} & \sqrt{2}[1-(-1)^n]\\
		\sqrt{2}[1-(-1)^n] & \sqrt{2}[1-(-1)^n] & {2}[1+(-1)^n]
	\end{pmatrix}.
	\label{}
\end{equation}
The prefactor then evaluates to $\left|2\cos\frac{\pi L}{8}\right|$, showing the same $L$ dependence as the lattice model calculation.

\twocolumngrid

\bibliography{SET}
\end{document}